\documentclass[11pt,a4paper]{article}

\usepackage[utf8]{inputenc}
\usepackage[margin=2.5cm]{geometry}
\usepackage{amsmath,amssymb,amsthm,mathtools}
\usepackage{bm}
\usepackage{booktabs,array,longtable}
\usepackage{float}
\usepackage{xcolor}
\usepackage{graphicx}
\usepackage{enumitem}
\usepackage{microtype}
\usepackage[numbers,sort&compress]{natbib}
\usepackage{listings}
\usepackage{tikz}
\usetikzlibrary{arrows.meta,positioning,shapes.geometric,fit,calc}
\usepackage[colorlinks=true,linkcolor=blue!50!black,citecolor=green!40!black,urlcolor=blue!50!black]{hyperref}
\usepackage[capitalise,noabbrev]{cleveref}

\definecolor{codebg}{gray}{0.96}
\definecolor{codekw}{rgb}{0.1,0.2,0.6}
\definecolor{codestr}{rgb}{0.55,0.1,0.1}
\definecolor{codecom}{rgb}{0.25,0.45,0.25}
\lstdefinestyle{py}{
  language=Python, basicstyle=\ttfamily\footnotesize, backgroundcolor=\color{codebg},
  keywordstyle=\color{codekw}\bfseries, stringstyle=\color{codestr}, commentstyle=\color{codecom}\itshape,
  frame=single, rulecolor=\color{black!20}, breaklines=true, breakatwhitespace=false,
  columns=fullflexible, keepspaces=true, breakautoindent=true, postbreak=\mbox{\textcolor{gray}{$\hookrightarrow$}\space}, showstringspaces=false, xleftmargin=3pt, xrightmargin=3pt,
  aboveskip=6pt, belowskip=6pt, literate={\ \ }{{\ }}1}
\newcommand{\code}[1]{\texttt{#1}}
\newcommand{\fs}{\textsc{fieldsmith}}

\newcommand{\dd}{\mathrm{d}}
\newcommand{\ii}{\mathrm{i}}
\newcommand{\ee}{\mathrm{e}}
\newcommand{\pd}{\partial}
\newcommand{\tr}{\operatorname{tr}}
\newcommand{\EL}[1]{\frac{\delta S}{\delta #1}}
\newcommand{\dual}[1]{\widetilde{#1}}
\renewcommand{\Box}{\square}
\newcommand{\vect}[1]{\bm{#1}}
\newcommand{\order}[1]{\mathcal{O}(#1)}
\newcommand{\weq}{\approx}          
\newcommand{\PB}[2]{\{#1,#2\}}
\DeclareMathOperator{\sgn}{sgn}

\theoremstyle{definition}

\theoremstyle{plain}

\newcounter{ex}
\crefname{ex}{Example}{Examples}
\Crefname{ex}{Example}{Examples}
\newcommand{\exheading}[4]{%
  \refstepcounter{ex}\label{#1}%
  \subsubsection*{Example~\theex: #2}%
  \addcontentsline{toc}{subsubsection}{Example~\theex: #2}%
  \noindent{\small\textit{Level: #4.\ Source: \texttt{examples/#3}.}}\par\smallskip}

\title{\textbf{\fs{}: Automated symbolic derivation in classical field theory and gauge theory from \LaTeX{} input}} 

\author{Nikola Mirkov \\    
\small `Vin\v{c}a' Institute of Nuclear Sciences--National Institute of the Republic of Serbia,\\ \small University of Belgrade, Mike Petrovi\'{c}a Alasa 12-14, 11351 Belgrade, Serbia\\
\small \texttt{nmirkov@vin.bg.ac.rs}}
\date{}

\begin{document}
\maketitle

\begin{abstract}
\noindent
We present \fs{}, an open-source Python library for the automated derivation of analytical results in classical field theory, and non-abelian gauge theories. A Lagrangian is entered the way it is written on a blackboard, in \LaTeX{} index notation, and the library produces and \emph{verifies} Euler--Lagrange equations (of arbitrary derivative order, on flat or curved backgrounds), Noether currents and identities, canonical and Hilbert energy--momentum tensors, canonical momenta and Hamiltonians, local Poisson brackets and a Dirac--Bergmann constraint analysis with degree-of-freedom counting, linearised wave equations and dispersion relations, gauge-covariant derivatives and field strengths for products of unitary groups, and mass spectra after spontaneous symmetry breaking. A fitting routine translates component-level output back into covariant index notation, and a grading module checks students' submitted field equations, identities and conserved currents. The implementation is a compact (about 2,700 lines) layer on top of \textsc{SymPy}: tensors are stored by components with labelled, variance-aware slots, and every statement is established by exact symbolic computation rather than pattern matching.
This document is the companion guide to the library. It explains the intent and design, describes the algorithms, and then works through eighteen self-checking examples, ranging from the Klein--Gordon field to the BPST instanton and the 't~Hooft--Polyakov monopole, each preceded by the background physics needed to follow it. The examples double as a compact introduction to classical field theory and gauge theory as seen through a computational lens. We are explicit about the limitations of the approach.
\end{abstract}

\tableofcontents
\clearpage

\section{Introduction}
\label{sec:intro}

Classical field theory is the language in which most of modern fundamental physics is written. Electromagnetism, the Klein--Gordon and Dirac fields, Yang--Mills theory, the Higgs mechanism, topological solitons and the coupling of matter to gravity are all introduced, in the first instance, as \emph{classical} field theories: a Lagrangian density, an action principle, and a set of consequences obtained by calculation. The calculations are routine in principle and tedious in practice. Varying a Yang--Mills action with matter, verifying that a given Noether current is conserved, counting the physical degrees of freedom of a massive vector field, or checking that a proposed instanton satisfies the field equations involves long chains of index manipulation in which a single sign error silently invalidates everything downstream.

Computer algebra has long been used for such work, but the tools that exist target neighbouring problems. General-purpose computer algebra systems provide the calculus but not the field-theoretic structure. Dedicated tensor systems such as \textsc{xAct} for \textsc{Mathematica} and \textsc{Cadabra} \cite{Peeters2007} are powerful for canonicalising tensor polynomials and for general relativity; \textsc{FORM} \cite{Vermaseren2000} and \textsc{FeynCalc} \cite{Mertig1991,Shtabovenko2016} are built for the algebra of perturbative amplitudes. What we found missing is a tool that sits at the level at which a graduate course in classical field theory is taught: a student or teacher writes down a Lagrangian, asks the questions that the course asks of it, and gets answers that are \emph{checked}. The present work describes such a tool.

\subsection{The \fs{} workflow}

\fs{} is a Python library, built on \textsc{SymPy} \cite{Meurer2017}, with the following workflow.
\begin{enumerate}[nosep]
\item The user declares fields, parameters and the spacetime, and writes the Lagrangian density as a \LaTeX{} string in index notation, e.g.\ \verb|-\frac14 F^a_{\mu\nu}F^{a\mu\nu}|.
\item A parser expands the string into explicit expressions in the coordinate components, summing over repeated indices, applying derivative and covariant-derivative operators, and inserting metric, Levi-Civita and Lie-algebra structure constants where required.
\item Analysis routines then operate on the resulting action: field equations, symmetries and conserved currents, energy--momentum tensors, canonical structure and constraints, linearised spectra, and gauge-theoretic data such as vector-boson mass matrices.
\item Because results are explicit expressions, \emph{any} claim can be tested: an identity written in \LaTeX{} is verified, a candidate solution is substituted into the field equations, and a component-level result is re-expressed in covariant index notation by solving a small linear problem.
\end{enumerate}
A minimal session already shows the flavour. For the electromagnetic field:
\begin{lstlisting}
from fieldsmith import Model
from fieldsmith.analysis import fit
from fieldsmith.hamiltonian import Hamiltonian

M = Model()
M.vector("A")
M.define(r"F_{\mu\nu}", r"\partial_\mu A_\nu - \partial_\nu A_\mu")
L = M.lagrangian(r"-\frac14 F_{\mu\nu}F^{\mu\nu}")
E = M.euler_lagrange(L)
print(fit(M, E.tensor("A"), {"dF": r"\partial_\mu F^{\mu\nu}"}))                  # {'dF': 1}
print(M.check(r"\partial_\nu\partial_\mu F^{\mu\nu} = 0"))                         # True
print(M.check(r"\partial_\mu (\frac12\epsilon^{\mu\nu\rho\sigma}F_{\rho\sigma}) = 0"))  # Bianchi
print(Hamiltonian(M, L).dirac_analysis().summary())
\end{lstlisting}
The output confirms that the field equation is $\pd_\mu F^{\mu\nu}=0$ (with unit coefficient), that $\pd_\nu\pd_\mu F^{\mu\nu}\equiv 0$ and that the dual field strength is divergence-free, and it reports two first-class constraints and two physical degrees of freedom per spatial point. Each statement has been established by exact symbolic computation on the components of $A_\mu$.

\subsection{Contributions and scope}

The main contributions are the following.
\begin{itemize}[nosep]
\item A \emph{LaTeX-first interface} to classical field theory: index notation with Einstein summation, derivative operators with explicit scoping rules, conjugation, accents, gamma matrices, covariant derivatives, and operator-valued expressions such as $(\ii\gamma^\mu D_\mu-m)\psi$ (\cref{sec:parser}).
\item A \emph{complete derivation pipeline} from action to spectrum: Euler--Lagrange equations of any derivative order; Noether currents for arbitrary continuous symmetries, including spacetime symmetries with boundary terms, together with the Noether identity as an independent check; canonical and Hilbert energy--momentum tensors; a Dirac--Bergmann analysis; and plane-wave dispersion analysis that handles gauge null directions (\cref{sec:framework,sec:impl}).
\item A \emph{gauge-theory module} for $\mathrm{SU}(N)$ and $\mathrm U(1)$ factors, matter multiplets and spinors, with field strengths, covariant derivatives on arbitrary representations, local gauge variations and mass matrices (\cref{sec:gauge-impl}).
\item A \emph{translation from components back to covariant notation} (\code{fit}), which makes the output readable and doubles as a decision procedure for statements of the form ``this tensor equals a combination of these covariant building blocks'' (\cref{sec:fit}).
\item A \emph{teaching toolkit} that grades submitted field equations, identities and currents, tolerant of equivalent forms but not of wrong ones (\cref{sec:grading}).
\item Eighteen \emph{self-checking worked examples} (\cref{part:examples}) covering the standard curriculum and several research-level topics. Each is both a documented derivation and a regression test.
\end{itemize}

The scope is deliberately \emph{classical}: fields are commuting functions on spacetime. Quantum questions enter only through classical precursors (degree-of-freedom counts, mass spectra, constraint structure). Section~\ref{sec:limits} lists what the library does not do.

\subsection{Organisation of this document}

\Cref{sec:intent} states the intent and design principles. \Cref{sec:usecases} describes the use cases we had in mind: teaching, assessment, exploration and research prototyping. \Cref{sec:framework} summarises the theoretical framework that the library implements, so that the document is self-contained. \Cref{sec:impl} describes the implementation in detail. \Cref{part:examples} then works through eighteen examples one by one. For each we give the physical background---which, taken together, forms a concise introduction to classical field and gauge theory---followed by the code, the output produced by the library, and an interpretation. \Cref{sec:discussion} discusses limitations, related software and future directions. The appendices collect conventions, the \LaTeX{} grammar, a compact API summary, and instructions for reproducing every number quoted here.

\subsection{Notation and conventions}

Unless stated otherwise we use natural units $\hbar=c=1$, the metric signature $(+,-,-,-)$, $\epsilon^{0123}=+1$, and Greek indices running over spacetime. The gauge conventions follow Peskin and Schroeder \cite{PeskinSchroeder}: $D_\mu=\pd_\mu-\ii g A_\mu^aT^a$ and $F^a_{\mu\nu}=\pd_\mu A_\nu^a-\pd_\nu A_\mu^a+gf^{abc}A_\mu^bA_\nu^c$. \Cref{app:conventions} lists everything, including the points at which the library's numbering differs from the physicist's (spacetime components are $0$-based; adjoint indices are $1$-based).

\section{Intent and design principles}
\label{sec:intent}

\fs{} was written for a specific purpose: to support the teaching of classical field theory and gauge theory to upper-level undergraduates and early graduate students, and to be a scaffold on which advanced homework and exploratory research calculations can be built. Several design decisions follow directly from that purpose.

\paragraph{Verify, do not pattern-match.}
A teaching tool that returns a plausible-looking but unchecked formula is worse than no tool. Whenever the library can establish a statement by exact computation, it does so, and every answer is an explicit expression that can be inspected. The Euler--Lagrange equations are \emph{derived} from the action, not looked up; the Bianchi identity is \emph{verified} by expanding both sides on components; the conservation of a Noether current is confirmed through the independent Noether identity (\cref{sec:noether}); a claimed solution is inserted into the equations. The library also contains negative controls: its fitting routine, for instance, returns \code{None} when a target is not in the span of the proposed building blocks, and the test-suite checks this behaviour explicitly.

\paragraph{Write physics, not code.}
The input language is \LaTeX{} in index notation because that is what physicists write. A student who can type $\frac12\pd_\mu\phi\,\pd^\mu\phi-\frac12m^2\phi^2$ in a problem set can type the same string to the library. This lowers the barrier to using the tool for homework and makes the examples readable as physics.

\paragraph{Components inside, covariant outside.}
Internally, fields are SymPy functions of the coordinates and tensors are stored by their components. This choice trades efficiency for transparency and robustness: there is no abstract-index canonicaliser that can be wrong, every intermediate quantity can be printed and plotted, and any result can be handed to the rest of SymPy. The price is that the cost grows with spacetime dimension and group size (\cref{sec:limits}). To compensate for the loss of readability, the \code{fit} routine of \cref{sec:fit} maps component results back to covariant form.

\paragraph{A small, readable core.}
The entire library is about $2{,}700$ lines of Python. A graduate student can read the Noether routine in an afternoon and see exactly how the current is built. We regard this as a feature: a teaching library should be inspectable, and a research prototype should be hackable.

\paragraph{Explicit limits.}
We prefer an explicit \code{NotImplementedError} to a wrong answer. The canonical module refuses Lagrangians with higher time derivatives or non-quadratic velocity dependence; the Hilbert-tensor routine refuses curved backgrounds. The limitations are listed in \cref{sec:limits} and repeated in the documentation shipped with the code.

\paragraph{Examples and tests are combined.}
Each of the eighteen examples ends in explicit pass/fail assertions on known results (\cref{sec:testing}). They therefore serve three purposes at once: documentation for the user, a regression suite for the developer, and a catalogue of results with which one can compare an unfamiliar tool.

\subsection{The current limits of the library}

\fs{} is not a Feynman-diagram generator, a quantum field theory package, or a numerical relativity code. It does not replace analytic insight: it removes the algebraic burden so that the user can concentrate on what the results mean. Likewise it is not a general tensor-algebra system; there is no canonicaliser for abstract index expressions and no support for Riemann-tensor monomials in arbitrary dimension. Finally, it is not yet a tool for gravitational dynamics: curvature tensors and covariant derivatives are available (\cref{sec:curved}), but the metric is a background, not a dynamical field.

\section{Use cases}
\label{sec:usecases}

We describe the scenarios the library was designed for. Each corresponds to a concrete workflow, and most are illustrated by one or more of the examples in \cref{part:examples}.

\subsection{Teaching: derivations on demand}

In lectures, the point of a derivation is usually its \emph{structure}, but the algebra is where time is lost. With \fs{} the instructor can write a Lagrangian, obtain the field equations, and move on to the physics. Typical live demonstrations include: the Klein--Gordon equation and its plane-wave dispersion (\cref{ex:kg}), the Maxwell equations and the Bianchi identity (\cref{ex:maxwell}), the Proca equation and the implied Lorenz condition (\cref{ex:proca}), the Higgs mechanism with explicit mass formulas (\cref{ex:higgs,ex:ew}), and Noether's theorem for translations, boosts and rotations (\cref{ex:symmetries}). Because results come back as explicit expressions, they can be compared directly with the textbook formulas and the audience can see the match.

\Cref{tab:course} maps a standard course outline to the library calls used. The accompanying documentation includes a suggested twelve-week syllabus with exercises.
\begin{table}[ht]
\centering\small
\begin{tabular}{@{}p{4.6cm}p{5.3cm}p{4.6cm}@{}}
\toprule
Course topic & Library call(s) & Example \\
\midrule
Action principle, field equations & \code{euler\_lagrange}, \code{check\_solution} & \ref{ex:kg}, \ref{ex:kink} \\
Solitons, Bogomolny bounds & \code{check\_solution}, energy density & \ref{ex:kink}, \ref{ex:bpst}, \ref{ex:monopole} \\
Symmetries and conserved currents & \code{noether\_current}, \code{symmetry\_defect} & \ref{ex:symmetries}, \ref{ex:goldstone} \\
Energy--momentum tensors & \code{canonical\_stress\_tensor}, \code{hilbert\_stress\_tensor} & \ref{ex:stress} \\
Gauge invariance, constraints & \code{gauge\_variation}, \code{dirac\_analysis} & \ref{ex:maxwell}, \ref{ex:proca} \\
Spontaneous symmetry breaking & \code{scalar\_mass\_matrix}, \code{gauge\_mass\_matrix} & \ref{ex:higgs}, \ref{ex:goldstone}, \ref{ex:ew} \\
Yang--Mills theory & \code{GaugeTheory}, \code{LieAlgebra} & \ref{ex:ym} \\
Spinors & \code{add\_matter(...,spinor=True)} & \ref{ex:qed} \\
Fields on curved backgrounds & \code{Spacetime}, \verb|\nabla| & \ref{ex:frw} \\
Wave propagation, dispersion & \code{linearize}, \code{dispersion} & \ref{ex:kg}, \ref{ex:axion}, \ref{ex:eh} \\
\bottomrule
\end{tabular}
\caption{Mapping a typical classical-field-theory course onto the library.}
\label{tab:course}
\end{table}

\subsection{Homework and assessment}

\paragraph{Designing problems.} A teacher can pose problems with guaranteed-correct answer keys: ``derive the field equation of $\mathcal L=\dots$'', ``show that $\delta\phi=\dots$ is a symmetry and find the current'', ``count the degrees of freedom of $\dots$'', ``find the mass spectrum after $\langle\Phi\rangle=\dots$''. Variants of a problem are obtained by changing the Lagrangian and rerunning the script. Because the cost of producing an answer key is a few lines of code, it becomes practical to give each student a different variant.

\paragraph{Automatic grading.} The \code{teaching} module compares a submitted answer, as a \LaTeX{} string, with the correct one. Equivalent forms are accepted (reordered terms, relabelled dummy indices, overall sign or normalisation for equations of motion), while a wrong sign in a mass term or a forgotten term is detected and reported with a message. Details are in \cref{sec:grading}. Because the student's input is parsed, never executed, grading is safe to run on untrusted submissions.

\subsection{Checking hand calculations and published formulas}

A researcher deriving a new result by hand can use the library as an independent check at each step: Is this field equation consistent with that Lagrangian? Is this current really conserved? Does this ansatz solve the equations? The \code{check} and \code{check\_solution} interfaces are intended for this use, and the \code{fit} routine reports whether a messy expression is a combination of the covariant quantities the researcher expects. The same ability is useful when reading a paper: type in its Lagrangian and test its claims.

\subsection{Exploring modified and effective theories}

Because the machinery does not assume a particular Lagrangian, it is straightforward to explore variants: add a term and ask what changes. The frontier examples (\cref{sec:frontier}) are written precisely so that they can be edited in this way. For instance, adding $-\tfrac g4aF\dual F$ to Maxwell theory yields axion electrodynamics and its birefringence (\cref{ex:axion}); adding $c_1(F^2)^2+c_2(F\dual F)^2$ gives the Euler--Heisenberg effective theory and the vacuum refractive indices (\cref{ex:eh}); adding $\tfrac{a^2}{2}(\pd_\mu F^{\mu\nu})^2$ gives Podolsky electrodynamics, a fourth-order theory with a finite Coulomb self-energy (\cref{ex:podolsky}).

\subsection{Research prototyping}

For new theories, the first questions are usually structural: how many degrees of freedom does it have, which constraints are first class (gauge symmetries), what is the dispersion relation of small perturbations about a background, and is there a conserved current for a given symmetry? The Dirac--Bergmann module and the dispersion module answer exactly these. We have used them to reproduce the textbook counts (Maxwell $2$, Proca $3$, Dirac $4$, topologically massive electrodynamics $1$), and the same functions can be pointed at an unfamiliar vector, tensor or higher-form theory as a first diagnostic. The result is not a proof of consistency---the algorithms are classical, restricted (\cref{sec:limits}), and not a substitute for a careful analysis---but it identifies pathologies early and cheaply.

\subsection{Self-study and cross-checking other software}

Finally, a self-learner can use the examples as worked solutions and modify them. The examples also provide benchmarks for other tools: each prints well-defined results (e.g.\ the $7{:}4$ ratio of the Euler--Heisenberg refractive indices, the instanton action $8\pi^2/g^2$, the monopole mass $4\pi v/g$) that another implementation can be compared against.

\section{Theoretical framework}
\label{sec:framework}

This section summarises the classical field theory that the library implements, in the form in which it is implemented. It is intended to make the paper self-contained; readers who know the material can skip to \cref{sec:impl}. The standard references are Landau and Lifshitz \cite{LandauLifshitz}, Weinberg \cite{WeinbergQFT1,WeinbergQFT2}, Peskin and Schroeder \cite{PeskinSchroeder} and, for constrained systems, Henneaux and Teitelboim \cite{HenneauxTeitelboim}.

\subsection{Action principle and field equations}
\label{sec:fw-el}

A classical field theory is specified by a set of fields $\varphi_I(x)$ on a $d$-dimensional spacetime with coordinates $x^\mu$ and a Lagrangian density $\mathcal L$, a function of the fields and finitely many of their derivatives. The action is
\begin{equation}
S[\varphi]=\int\dd^dx\,\sqrt{|g|}\;\mathcal L\big(\varphi_I,\pd_\mu\varphi_I,\pd_\mu\pd_\nu\varphi_I,\dots\big),
\label{eq:action}
\end{equation}
with $g=\det g_{\mu\nu}$ and $\sqrt{|g|}=1$ in Cartesian coordinates on Minkowski space. Stationarity under variations vanishing at the boundary gives the Euler--Lagrange (EL) equations
\begin{equation}
E_I\equiv\EL{\varphi_I}=\sum_{k\ge0}(-1)^k\,\pd_{\alpha_1}\!\cdots\pd_{\alpha_k}\frac{\pd\,(\sqrt{|g|}\mathcal L)}{\pd(\pd_{\alpha_1}\!\cdots\pd_{\alpha_k}\varphi_I)}\Big/\sqrt{|g|}=0,
\label{eq:EL}
\end{equation}
where the derivative with respect to a symmetrised multi-derivative is understood to be shared equally among its orderings. We call $E_I$ the \emph{equation-of-motion expression} of the field $\varphi_I$; it is the quantity that vanishes on shell, and its overall sign is fixed by the convention in \cref{eq:EL}. For complex fields we treat $\phi$ and $\phi^*$ as independent, so $\delta S/\delta\phi^*$ is the equation of motion of $\phi$. The presence of derivatives of order $k>1$ in \cref{eq:EL} is what permits higher-derivative theories such as Podolsky's (\cref{ex:podolsky}).

\subsection{Symmetries and Noether's theorem}
\label{sec:fw-noether}

Let $\varphi_I\to\varphi_I+\epsilon\,\delta\varphi_I$, with $\delta\varphi_I$ a local function of the fields and their derivatives. If the Lagrangian density changes by a total divergence, $\delta\mathcal L=\pd_\mu K^\mu$, the variation is a symmetry. For a first-order Lagrangian, the standard computation gives
\begin{equation}
\delta\mathcal L=\sum_IE_I\,\delta\varphi_I+\pd_\mu\Big(\sum_I\frac{\pd\mathcal L}{\pd(\pd_\mu\varphi_I)}\delta\varphi_I\Big).
\label{eq:dL}
\end{equation}
Defining $J^\mu=\sum_I\frac{\pd\mathcal L}{\pd(\pd_\mu\varphi_I)}\delta\varphi_I-K^\mu$, we obtain the \emph{Noether identity}
\begin{equation}
\pd_\mu J^\mu+\sum_IE_I\,\delta\varphi_I\equiv0,
\label{eq:noether-id}
\end{equation}
which holds \emph{off shell}, identically in the fields. On shell ($E_I=0$) the current is conserved, $\pd_\mu J^\mu=0$, and the charge $Q=\int\dd^{d-1}x\,J^0$ is constant in time. For Lagrangians depending on derivatives up to order $n$, \cref{eq:dL} acquires further integrations by parts and
\begin{equation}
J_0^\mu=\sum_{k=1}^{n}\sum_{j=0}^{k-1}(-1)^j\,\pd_{\nu_1}\!\!\cdots\pd_{\nu_j}\frac{\pd\mathcal L}{\pd(\pd_\mu\pd_{\nu_1}\!\cdots\pd_{\nu_j}\pd_{\beta_{j+1}}\!\cdots\pd_{\beta_k}\varphi_I)}\;\pd_{\beta_{j+1}}\!\!\cdots\pd_{\beta_k}\delta\varphi_I,\qquad J^\mu=J_0^\mu-K^\mu,
\label{eq:Jhigher}
\end{equation}
which is exactly the expression the library evaluates (\cref{sec:noether}); \cref{eq:noether-id} remains valid and is used as an independent check.

\paragraph{Spacetime symmetries.} For an infinitesimal coordinate change $x^\mu\to x^\mu+\epsilon\,\xi^\mu(x)$ under which the fields transform as scalars, the field variation is the transport term $\delta\varphi_I=-\xi^\mu\pd_\mu\varphi_I$ and the Lagrangian density transforms as $\delta\mathcal L=-\pd_\mu(\xi^\mu\mathcal L)$ when the symmetry is a symmetry of the action, i.e.\ $K^\mu=-\xi^\mu\mathcal L$. Translations $\xi^\mu=\delta^\mu_\nu$ give the canonical energy--momentum tensor, rotations and boosts give angular momentum and the boost charge, and the scale transformation $\xi^\mu=x^\mu$ with the appropriate field weights gives the dilatation current (\cref{ex:symmetries}).

\subsection{Energy--momentum tensors}
\label{sec:fw-T}

Two tensors play a role. The \emph{canonical} (Noether) tensor, from translations, is
\begin{equation}
T^\mu{}_\nu=\sum_I\frac{\pd\mathcal L}{\pd(\pd_\mu\varphi_I)}\pd_\nu\varphi_I-\delta^\mu_\nu\mathcal L,
\label{eq:Tcan}
\end{equation}
which is conserved on shell but in general neither symmetric nor gauge invariant. The \emph{Hilbert} (metric) tensor is obtained by varying the action with respect to the metric,
\begin{equation}
T_{\mu\nu}=\frac{2}{\sqrt{|g|}}\frac{\delta(\sqrt{|g|}\,\mathcal L)}{\delta g^{\mu\nu}}\qquad\text{(signature }(+,-,-,-)\text{)},
\label{eq:Thilb}
\end{equation}
is symmetric by construction, and is gauge invariant when $\mathcal L$ is. The two differ by the Belinfante improvement, a term of the form $\pd_\lambda B^{\lambda\mu\nu}$ with $B$ antisymmetric in its first two indices, plus terms proportional to the equations of motion (\cref{ex:stress}). The difference matters physically: only the Hilbert tensor sources gravity.

\subsection{Gauge symmetries, constraints and degrees of freedom}
\label{sec:fw-dirac}

A \emph{gauge} symmetry is a symmetry with parameters that are arbitrary functions of spacetime. By Noether's second theorem it implies identities among the field equations; for Maxwell theory $\pd_\nu(\pd_\mu F^{\mu\nu})\equiv0$ is the statement that the equation of motion does not determine all of $A_\mu$. In canonical language this appears as constraints.

Given a Lagrangian that is at most quadratic in the velocities $\dot\varphi_I$, the canonical momenta are $\pi^I=\pd\mathcal L/\pd\dot\varphi_I$ and, when the Hessian $\pd^2\mathcal L/\pd\dot\varphi_I\pd\dot\varphi_J$ is regular, the Hamiltonian density is $\mathcal H=\pi^I\dot\varphi_I-\mathcal L$. The Poisson bracket of functionals is
\begin{equation}
\PB FG=\int\dd^{d-1}x\sum_I\Big(\frac{\delta F}{\delta\varphi_I}\frac{\delta G}{\delta\pi^I}-\frac{\delta F}{\delta\pi^I}\frac{\delta G}{\delta\varphi_I}\Big),
\label{eq:PB}
\end{equation}
and functional derivatives involve spatial derivatives only. If the Hessian is singular, there are \emph{primary constraints} $\phi_a\weq0$ (here $\weq$ is Dirac's weak equality). The Dirac--Bergmann algorithm \cite{Dirac1950,HenneauxTeitelboim} demands that the constraints are preserved in time, $\dot\phi_a=\PB{\phi_a}{H_T}\weq0$ with $H_T=H_c+u^a\phi_a$. Each condition either determines a multiplier $u^a$, or produces a new (\emph{secondary}) constraint, or is trivially satisfied. The procedure is iterated until it closes. A constraint is \emph{first class} if its bracket with all constraints is weakly zero, and \emph{second class} otherwise. First-class constraints generate gauge transformations. If the phase space has $2N$ dimensions per spatial point, with $N_1$ first-class and $N_2$ second-class constraints, the number of physical degrees of freedom is
\begin{equation}
n_{\mathrm{dof}}=\tfrac12\big(2N-2N_1-N_2\big).
\label{eq:dof}
\end{equation}
This is the count that the library returns. For Maxwell theory, $N=4$, $N_1=2$, $N_2=0$, so $n_{\mathrm{dof}}=2$; for Proca, $N=4$, $N_1=0$, $N_2=2$ gives $3$ (\cref{ex:maxwell,ex:proca}).

\subsection{Linearised waves}
\label{sec:fw-waves}

To study small perturbations about a background $\bar\varphi_I$, write $\varphi_I=\bar\varphi_I+\epsilon\,\psi_I$ and expand the field equations to first order in $\epsilon$. For a homogeneous (constant-coefficient) linear system, the substitution of plane waves $\psi_I=a_I\,\ee^{-\ii(\omega t-\vect k\cdot\vect x)}$ gives a matrix equation $K_{IJ}(\omega,\vect k)\,a_J=0$. Non-trivial solutions require a vanishing determinant,
\begin{equation}
\det K(\omega,\vect k)=0,
\label{eq:disp}
\end{equation}
which is the dispersion relation. In a gauge theory $K$ has null vectors for every $(\omega,\vect k)$ (pure-gauge polarisations), so $\det K\equiv0$; the physical modes are then the points where the rank of $K$ drops below its generic value $r$. These are common zeros of all $r\times r$ minors, which we compute as the greatest common divisor of the minors (\cref{sec:waves-impl}).

\subsection{Gauge theory}
\label{sec:fw-gauge}

Let $T^a$ be Hermitian generators of a compact Lie algebra, $[T^a,T^b]=\ii f^{abc}T^c$, normalised for $\mathrm{SU}(N)$ by $\tr(T^aT^b)=\tfrac12\delta^{ab}$. A multiplet $\phi$ in a representation $R$ transforms as $\delta\phi=\ii\alpha^a(x)T^a_R\phi$. The covariant derivative and field strength
\begin{equation}
D_\mu=\pd_\mu-\ii gA_\mu^aT^a,\qquad F_{\mu\nu}^a=\pd_\mu A_\nu^a-\pd_\nu A_\mu^a+gf^{abc}A_\mu^bA_\nu^c,
\label{eq:DF}
\end{equation}
transform covariantly if $\delta A_\mu^a=g^{-1}\pd_\mu\alpha^a+f^{abc}A_\mu^b\alpha^c$. The curvature is the commutator, $[D_\mu,D_\nu]=-\ii gF_{\mu\nu}^aT^a$, from which the Bianchi identity $D_\mu\dual F^{a\,\mu\nu}=0$ follows. The Yang--Mills Lagrangian $-\tfrac14F^a_{\mu\nu}F^{a\,\mu\nu}$ gives $D_\mu F^{a\,\mu\nu}=0$ (\cref{ex:ym}). When a scalar multiplet acquires a vacuum expectation value $\langle\phi\rangle=v$ breaking the gauge group, the kinetic term $(D_\mu\phi)^\dagger D^\mu\phi$ contains $\tfrac12M^2_{ab}A^a_\mu A^{b\,\mu}$ with $M^2_{ab}=g^2\,v^\dagger\{T^a,T^b\}v$; its eigenvalues are the squared masses of the vector bosons and its null vectors the unbroken directions (\cref{ex:higgs,ex:ew}).

\section{Implementation}
\label{sec:impl}

\subsection{Architecture}
\label{sec:arch}

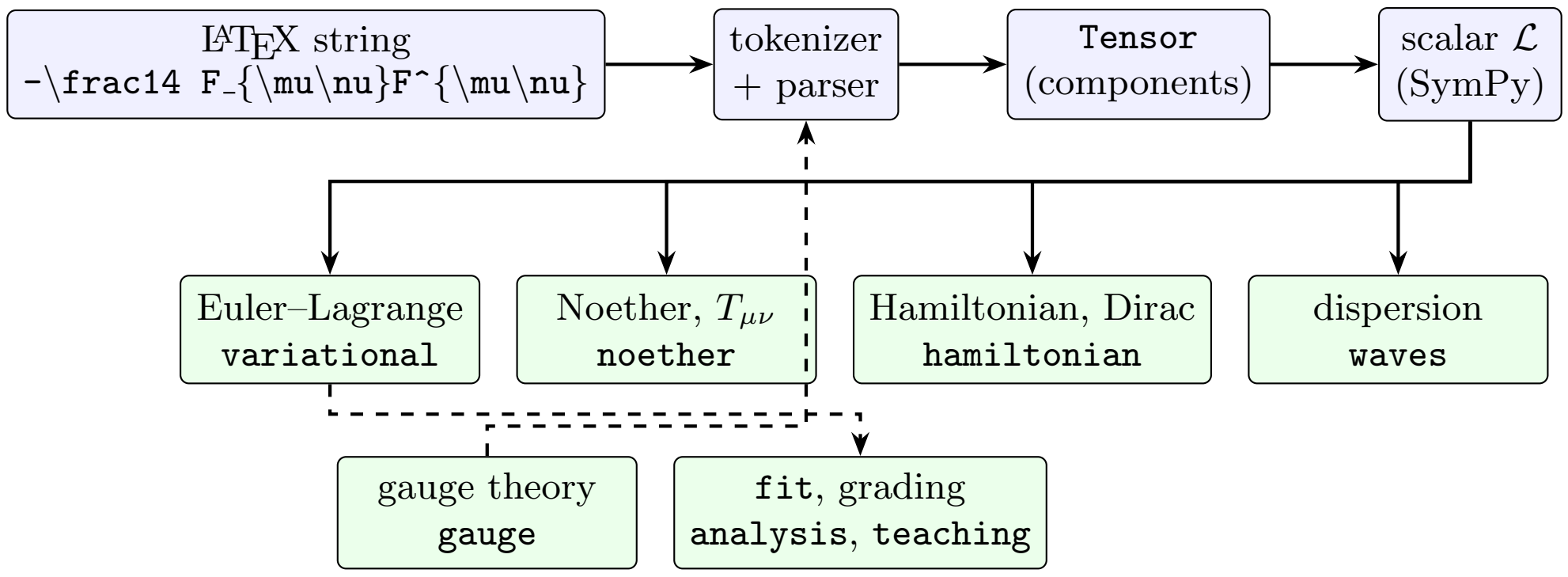
\begin{figure}[ht]
\centering
\begin{tikzpicture}[
  box/.style={draw,rounded corners=2pt,align=center,minimum height=0.9cm,font=\footnotesize,fill=blue!6},
  mod/.style={draw,rounded corners=2pt,align=center,minimum height=0.9cm,minimum width=2.5cm,font=\footnotesize,fill=green!8},
  arr/.style={-{Stealth[length=2mm]},thick}]
 \node[box] (tex) {\LaTeX{} string\\[-1pt]\texttt{-\textbackslash frac14 F\_\{\textbackslash mu\textbackslash nu\}F\^{}\{\textbackslash mu\textbackslash nu\}}};
 \node[box,right=0.9cm of tex] (parse) {tokenizer\\+ parser};
 \node[box,right=0.9cm of parse] (tens) {\code{Tensor}\\(components)};
 \node[box,right=0.9cm of tens] (L) {scalar $\mathcal L$\\(SymPy)};
 \node[mod,below=1.3cm of tex,xshift=0.2cm] (el) {Euler--Lagrange\\\code{variational}};
 \node[mod,right=0.3cm of el] (no) {Noether, $T_{\mu\nu}$\\\code{noether}};
 \node[mod,right=0.3cm of no] (ha) {Hamiltonian, Dirac\\\code{hamiltonian}};
 \node[mod,right=0.3cm of ha] (wa) {dispersion\\\code{waves}};
 \node[mod,below=0.6cm of no,xshift=-1.5cm] (ga) {gauge theory\\\code{gauge}};
 \node[mod,right=0.3cm of ga] (fi) {\code{fit}, grading\\\code{analysis}, \code{teaching}};
 \draw[arr] (tex)--(parse); \draw[arr] (parse)--(tens); \draw[arr] (tens)--(L);
 \draw[arr] (L.south)--++(0,-0.5)-|(el.north); \draw[arr] (L.south)--++(0,-0.5)-|(no.north);
 \draw[arr] (L.south)--++(0,-0.5)-|(ha.north); \draw[arr] (L.south)--++(0,-0.5)-|(wa.north);
 \draw[arr,dashed] (ga.north)--++(0,0.25)-|(parse.south);
 \draw[arr,dashed] (el.south)--++(0,-0.25)-|(fi.north);
\end{tikzpicture}
\caption{Data flow in \fs{}. Solid arrows: the main pipeline. Dashed arrows: the gauge module registers the covariant derivative \texttt{D} and the field strength with the parser; \code{fit} and the grader consume the output of the variational routines.}
\label{fig:arch}
\end{figure}

The library is organised as a pipeline (\cref{fig:arch}). A \code{Model} object owns the spacetime, the registries of fields, parameters and derived tensors, and the parser context. Strings pass through the tokenizer and parser to become \code{Tensor} objects (component dictionaries), which are scalars once all indices are contracted. The analysis modules take a scalar Lagrangian and return \code{EquationSet}, \code{NoetherCurrent}, \code{Hamiltonian} and similar objects, all of which carry explicit SymPy expressions.

\Cref{tab:modules} lists the modules and their sizes. The sizes are in physical lines of Python including docstrings.
\begin{table}[ht]
\centering\small
\begin{tabular}{@{}llr p{8cm}@{}}
\toprule
Module & Role & Lines & Contents \\
\midrule
\code{spacetime} & geometry & 119 & metrics; Christoffel, Riemann, Ricci, Einstein tensors; Minkowski/Euclidean factories \\
\code{tensors} & algebra & 372 & \code{Slot}, \code{Tensor}: contraction, raising/lowering, derivatives, symmetry tests \\
\code{latex} & parser & 547 & tokenizer, recursive-descent parser, operator scoping \\
\code{model} & registry & 399 & fields, parameters, \code{define}, \code{lagrangian}, \code{check} \\
\code{variational} & EL & 109 & \code{EquationSet}, \code{euler\_lagrange}, solution checking \\
\code{noether} & symmetries & 224 & Noether currents, stress tensors, complex variations \\
\code{hamiltonian} & canonical & 206 & momenta, $\mathcal H$, brackets, Dirac--Bergmann \\
\code{waves} & spectra & 107 & linearisation, plane waves, dispersion \\
\code{gauge} & gauge theory & 374 & \code{LieAlgebra}, \code{GaugeTheory}, mass matrices \\
\code{analysis} & utilities & 100 & \code{fit}, numerical residuals, scalar mass matrix \\
\code{teaching} & grading & 85 & field-equation, identity and current graders \\
\code{printing} & output & 40 & \LaTeX{} rendering of results \\
\bottomrule
\end{tabular}
\caption{Modules of \fs{}.}
\label{tab:modules}
\end{table}

\subsection{Spacetime and curved backgrounds}
\label{sec:curved}

A \code{Spacetime} holds the coordinates, the metric $g_{\mu\nu}$ and its inverse, $\sqrt{|g|}$ and a signature tag. Factories build $d$-dimensional Minkowski space (mostly-minus or mostly-plus) and Euclidean space; an arbitrary metric is supplied as a SymPy matrix. From it the module computes the Christoffel symbols $\Gamma^\lambda_{\mu\nu}=\tfrac12g^{\lambda\sigma}(\pd_\mu g_{\sigma\nu}+\pd_\nu g_{\sigma\mu}-\pd_\sigma g_{\mu\nu})$, the Riemann tensor (MTW convention \cite{MTW}), the Ricci tensor and scalar and the Einstein tensor. The covariant derivative \verb|\nabla_\mu| acts on tensors slot by slot, adding one Christoffel term per index with the sign dictated by its variance. The action density in \cref{eq:action} carries $\sqrt{|g|}$ automatically, so \code{euler\_lagrange} on a curved background returns the covariant equation of motion (e.g.\ with Hubble friction in FRW, \cref{ex:frw}). Curvature and covariant derivatives are tested on FRW and Schwarzschild; the metric is a fixed background.

\subsection{Tensors}
\label{sec:tensors}

A \code{Tensor} stores a list of slots and a dictionary mapping index tuples to component values. Each \code{Slot} records
\begin{itemize}[nosep]
\item a \emph{label} (\texttt{mu}, \texttt{a}, \texttt{i}, \dots),
\item a \emph{kind}: spacetime (\texttt{st}), spatial (\texttt{sp}; ranges over $1,\dots,d-1$ and is used for non-covariant 3-vector notation such as $E_i$ and $\epsilon_{ijk}$), adjoint of a gauge factor (\texttt{ad\_X}, ranging over the dimension of the algebra), or a user-defined family (flavour indices, for example),
\item a \emph{variance}: upper, lower, or none (adjoint and spatial indices),
\item the \emph{range} of values.
\end{itemize}
Component values are SymPy expressions or, for fields in a non-trivial representation, SymPy column matrices (a $\mathrm{SU}(2)$ doublet is a $2\times1$ matrix; a Dirac spinor a $4\times1$ matrix). Matrices multiply in the order written, and $1\times1$ matrices collapse to scalars, so that $\Phi^\dagger T^a\Phi$ is an ordinary number.

\paragraph{Contraction.} Multiplication of two tensors forms the outer product and then contracts every repeated label. A repeated spacetime label with opposite variance is a metric-free sum $\sum_\mu T_\mu S^\mu$; a spatial or adjoint label sums plainly. If the same spacetime label appears twice in the same position (e.g.\ $A_\mu A_\mu$), the default behaviour is to raise one copy with the metric, which is convenient but hides typographical errors; \code{Model(strict=True)} turns this into an error.

\paragraph{Metric operations.} \code{raise\_}, \code{lower} and \code{set\_var} apply $g^{\mu\nu}$ or $g_{\mu\nu}$ to a slot. Slots restricted to a sub-range (e.g.\ spatial indices on a spacetime axis) refuse the operation when the metric mixes the restricted and unrestricted directions.

\paragraph{Derivatives.} \code{partial} adds a new leading slot $\pd_\mu T$ obtained by differentiating each component; \code{nabla} adds Christoffel corrections on every spacetime slot; \code{dagger} conjugates the components and, for matrix-valued data, transposes them. Symmetry tests (\code{is\_symmetric}, \code{is\_antisymmetric}) compare a tensor with its transpose in a chosen pair of slots. (An earlier version of two examples tested symmetry by permuting labels in a way that reduced to the identity, so the assertion held vacuously; the dedicated transpose predicates were added to prevent this, and the examples now assert both a positive and a negative symmetry property.)

\sloppy\subsection{The \LaTeX{} parser}
\label{sec:parser}

The parser is a hand-written tokenizer and recursive-descent parser for a subset of \LaTeX{} (the grammar is in \cref{app:grammar}). The principal design question is how to treat derivative operators, because in \LaTeX{} there is no syntactic marker for where the argument of $\pd_\mu$ ends.

\paragraph{Scoping rule.} Operators act on the next postfix term. Thus
\begin{center}
\verb|\partial_\mu \phi^2|\quad means\quad $\pd_\mu(\phi^2)$,\qquad
\verb|\partial_\mu\phi\,\partial^\mu\phi|\quad means\quad $(\pd_\mu\phi)(\pd^\mu\phi)$,
\end{center}
and parentheses extend the scope: \verb|\partial_\mu (\phi\psi)|. The same rule governs \verb|\nabla_\mu|, \verb|\Box|, the gauge-covariant \verb|D_\mu| (registered when a gauge group is added) and the Dirac bar \verb|\bar|. Repeated derivatives are contracted after application, so \verb|\partial_\mu\partial^\mu A^\nu| is $\Box A^\nu$.

\paragraph{Operator-valued groups.} Expressions such as \verb|(i\gamma^\mu D_\mu - m)\psi| contain a differential operator inside parentheses. The parser represents the parenthesised group as an \code{OpExpr}, a deferred expression that is applied to the operand that follows. This is what makes the Dirac Lagrangian \verb|\bar\psi (i\gamma^\mu D_\mu - m)\psi| parse.

\paragraph{Symbols, subscripts, accents.} The parser must decide whether a subscript is an index or part of a name. A symbol such as \verb|m_a| or \verb|c_1| is a parameter if registered; \verb|g'| and \verb|g_s| are looked up as primed or subscripted couplings. A parameter named \texttt{g} shadows the built-in metric tensor $g_{\mu\nu}$ unless the symbol is followed by index-like subscripts. Accents are merged into the symbol name: \verb|\tilde{F}| becomes the symbol \texttt{Ftilde}, which the user defines like any other tensor (e.g.\ the dual field strength). Conjugation is written \verb|\phi^*|, \verb|\phi^\dagger| or \verb|(\ldots)^\dagger|.

\paragraph{Built-in tensors.} The metric $g_{\mu\nu}$ and $\eta_{\mu\nu}$, the Kronecker $\delta^\mu_\nu$ and the Levi-Civita symbol $\epsilon$ of any rank equal to the dimension are available. The latter is a tensor with $\epsilon^{0\cdots d-1}=+1/\sqrt{|g|}$; spatial $\epsilon_{ijk}$ (rank $d-1$ on the spatial slots) is used for 3-vector expressions such as the magnetic field. Gamma matrices are installed in the Dirac representation when a spinor is declared.

\paragraph{Other conveniences.} Fractions with the shorthand \verb|\frac12|, factorials (so that $\phi^4/4!$ is written naturally), elementary functions, the self-contraction shorthand \verb|(F_{\mu\nu})^2|, and spacing commands are all handled. \verb|\sum| and \verb|\int| are not supported: sums are written by repeated indices, and spacetime integrals are never written, because the action is implicit.

\subsection{The \code{Model} object}
\label{sec:model}

A \code{Model} keeps the following registries.
\begin{description}[nosep,leftmargin=1.2em]
\item[Parameters] declared with assumptions (\code{M.parameters("m lambda", positive=True)}); returned as SymPy symbols.
\item[Fields] \code{scalar}, \code{vector}, \code{tensor}, \code{multiplet}, general \code{field(name, index spaces, variances, \dots)}. A field with spacetime rank $r$ is stored as a tensor whose components are SymPy undefined functions $A_\mu\to\texttt{A[mu](t,x,y,z)}$, named by component; a complex field also registers its conjugate (named with a trailing star).
\item[Index spaces] \code{add\_index\_space} introduces a new family of labels with a range, used e.g.\ for the $N$ components of an $\mathrm{O}(N)$ vector (\cref{ex:goldstone}).
\item[Derived tensors] \code{define(lhs, rhs)} parses the right-hand side and stores the resulting tensor under the name on the left, with the stated index structure. Because definitions are evaluated eagerly, they can be used in subsequent strings exactly as symbols; for instance defining \verb|F_{\mu\nu}| as \verb|\partial_\mu A_\nu - \partial_\nu A_\mu| allows \verb|F^{\mu\nu}| later, with the metric applied.
\item[Operators and conjugation map] used by the parser for \verb|\bar|, \verb|\dagger| and the covariant derivative.
\end{description}
\code{lagrangian(s)} parses a string to a scalar and fully expands it. \code{check(s)} parses a string of the form \texttt{lhs = rhs}, brings both sides to common index structure and returns whether their difference vanishes after simplification; it is the engine for verifying identities (Bianchi, current conservation, algebraic relations).

\subsection{Euler--Lagrange equations}
\label{sec:el-impl}

\code{euler\_lagrange(L)} multiplies $\mathcal L$ by $\sqrt{|g|}$, calls SymPy's \code{euler\_equations} for each field function (this handles derivatives of arbitrary order), and divides the result by $\sqrt{|g|}$ so that the output is the covariant equation \cref{eq:EL}. The result is an \code{EquationSet}, indexable by position, by field-function or by name; \code{tensor(name)} assembles the equations of a vector or tensor field into a \code{Tensor} with the opposite variance ($E^\nu=\delta S/\delta A_\nu$), which is what \code{fit} consumes. \code{check\_solution} substitutes a mapping from field functions to expressions and tests for vanishing, either by symbolic simplification or, for transcendental solutions (breathers, monopole profiles), by evaluating the residual at random rational points to $30$ digits (\code{numeric=True}); free parameters are given random values and the residual is reported.

\subsection{Noether currents and energy--momentum tensors}
\label{sec:noether}

\paragraph{Algorithm.} Given $\mathcal L$ and a dictionary of field variations $\delta\varphi_I$ (for complex fields both the field and its conjugate are given, by a helper that conjugates the transformation), \code{noether\_current} proceeds as follows.
\begin{enumerate}[nosep]
\item Compute $\delta\mathcal L$ by substituting $\varphi_I\to\varphi_I+\epsilon\,\delta\varphi_I$ and differentiating with respect to $\epsilon$ at $\epsilon=0$.
\item If a boundary term $K^\mu$ is supplied, test that $\delta\mathcal L-\pd_\mu K^\mu$ vanishes identically; if not, raise an error (the \emph{symmetry check}). The defect is available separately as \code{symmetry\_defect}, and is how the library reports \emph{how} a symmetry is broken, e.g.\ $m^2\phi^2$ for the dilatation current of a massive field.
\item Collect the momenta $P^{\alpha_1\ldots\alpha_k}=\pd\mathcal L/\pd(\pd_{\alpha_1\ldots\alpha_k}\varphi_I)$, distributing the derivative of $\mathcal L$ with respect to a mixed partial equally over its distinct orderings, and assemble $J_0^\mu$ from \cref{eq:Jhigher} and $J^\mu=J_0^\mu-K^\mu$.
\item Verify the Noether identity \cref{eq:noether-id} with the equations from \code{euler\_lagrange}. This check is independent of the construction in step 3 and is what \code{NoetherCurrent.identity\_holds} runs.
\end{enumerate}
The canonical stress tensor $T^\mu{}_\nu$ is obtained by running the algorithm for $\delta\varphi_I=\pd_\nu\varphi_I$ with $K^\mu=\delta^\mu_\nu\mathcal L$ for each $\nu$, so it inherits the generality (higher derivatives) and the checking.

\paragraph{Hilbert tensor.} \code{hilbert\_stress\_tensor} implements \cref{eq:Thilb} by a first-order perturbation of the inverse metric, $g^{\mu\nu}=\eta^{\mu\nu}+h^{\mu\nu}$, with $g_{\mu\nu}=\eta_{\mu\nu}-\eta_{\mu\alpha}h^{\alpha\beta}\eta_{\beta\nu}$ and $\sqrt{|g|}=1-\tfrac12\tr(h\eta)$ to the same order. The Lagrangian string is \emph{re-parsed} in a model whose spacetime carries the perturbed metric, so every contraction picks up its dependence on $h$; the derivative of $\sqrt{|g|}\mathcal L$ with respect to $h^{\mu\nu}$ at $h=0$ gives $T_{\mu\nu}$ (with the factor $\tfrac12$ for off-diagonal entries and the sign appropriate to the signature). Because it re-parses, the function takes the Lagrangian as a \emph{string}, and any derived tensors it uses should be built from metric-independent pieces such as $F_{\mu\nu}=\pd_\mu A_\nu-\pd_\nu A_\mu$. It is restricted to flat backgrounds and bosonic fields (a spinor requires a tetrad, which is not implemented).

\subsection{Hamiltonian analysis}
\label{sec:ham-impl}

\paragraph{Construction.} The \code{Hamiltonian} class first rejects Lagrangians containing time derivatives of order higher than one (\code{NotImplementedError}: Ostrogradsky's construction \cite{Ostrogradsky1850,Woodard2007} is not implemented). It replaces $\dot\varphi_I$ by symbols $v_I$ and computes $p_I(\varphi,\pd_i\varphi,v)=\pd\mathcal L/\pd v_I$, building the matrix $W_{IJ}=\pd p_I/\pd v_J$. Fields with an identically zero row are \emph{singular}: their momenta give primary constraints $\pi^I-p_I|_{v=0}\weq0$ (this is how $\pi^0\weq0$ appears for $A_0$). The regular block is inverted by solving $\pi^I=p_I(v)$ for the velocities, and the canonical Hamiltonian density is $\mathcal H=\sum\pi^Iv_I-\mathcal L$ evaluated on the solution. The code refuses a velocity-dependent Hessian (e.g.\ Born--Infeld) because the inversion is then not algebraic in general.

\paragraph{Brackets.} Functional derivatives $\delta F/\delta\varphi$ are taken with spatial integration by parts only (time derivatives are kept as independent). The bracket \code{bracket(F, G)} returns the integrand of \cref{eq:PB}; two densities have a vanishing bracket if the integrand is a total spatial divergence, which \code{is\_divergence} detects by testing that all functional derivatives of the integrand vanish. Constraints are \emph{smeared} with arbitrary test functions $\xi(x),\eta(x)$ so that brackets of local densities are meaningful.

\paragraph{Dirac--Bergmann.} \code{dirac\_analysis} iterates the algorithm of \cref{sec:fw-dirac}. For each constraint $\phi_a$ it evaluates $\{\phi_a,H_c\}$ on the primary-constraint surface; if $\phi_a$ fails to commute with some primary constraint the condition fixes a multiplier and no new constraint arises, otherwise a non-vanishing remainder (after solving earlier constraints for a pivot momentum derivative) becomes a new secondary constraint. Once the list is complete, the bracket matrix between all constraints is computed on the constraint surface; a constraint is \emph{second class} if some bracket with it is non-zero (modulo total derivatives and the pivot reduction) and \emph{first class} otherwise. The degrees of freedom follow from \cref{eq:dof}. The algorithm is bounded by a maximum number of rounds (default three), which covers all textbook theories we tested (up to tertiary constraints).

\subsection{Waves and dispersion relations}
\label{sec:waves-impl}

\code{linearize(model, E, background)} substitutes $\varphi_I=\bar\varphi_I+\epsilon\,\psi_I$ into the equation-of-motion expressions, expands to $\order\epsilon$ and returns the linear expressions together with the perturbation functions $\psi_I$ (named \texttt{d\_<field>}). The background may depend on position, e.g.\ $\bar A_y=B_0x$ for a constant magnetic field (\cref{ex:eh}), provided the coefficients in the linearised equations are constant at the wave-vector of interest. \code{PlaneWave} substitutes $\psi_I\to a_I\ee^{-\ii(\omega t-\vect k\cdot\vect x)}$ and reads off the matrix $K$ column by column; in Euclidean signature the phase is $\ee^{\ii\vect k\cdot\vect x}$. \code{dispersion} returns $K$, its determinant (factored), the roots in $\omega$ and the rank. If $\det K\equiv0$ and the rank is $r>0$, it forms all $r\times r$ minors of $K$, factors them and takes their \emph{greatest common divisor}: this is the polynomial whose zero locus is exactly where the rank falls below its generic value, i.e.\ the physical dispersion relation (\code{pdet}). An earlier implementation used the \emph{sum} of the minors and produced spurious roots; the gcd resolves this. The method captures the divisorial part of the rank-drop locus, which suffices for the theories we treat but is not a complete characterisation in general.

\subsection{Gauge theory}
\label{sec:gauge-impl}

\paragraph{Lie algebras.} \code{LieAlgebra.su(N)} builds the generalised Gell-Mann matrices, normalised to $\tr(T^aT^b)=\tfrac12\delta^{ab}$, and computes the structure constants $f^{abc}=-2\ii\,\tr([T^a,T^b]T^c)$ exactly; \code{u1()} gives the one-dimensional abelian algebra. The class exposes the Jacobi identity check, the adjoint matrices $(T^a_{\mathrm{adj}})_{bc}=-\ii f^{abc}$ and the quadratic Casimir ($C_F=(N^2-1)/2N$: $3/4$ for $\mathrm{SU}(2)$, $4/3$ for $\mathrm{SU}(3)$).

\paragraph{\code{GaugeTheory}.} \code{add\_group(name, algebra, coupling, strength=...)} registers a gauge field $A^a_\mu$ with adjoint index family, a coupling constant (a symbol, e.g.\ \texttt{g} or \texttt{g'}), and the field strength tensor of \cref{eq:DF} under a name (default \texttt{F}) that becomes available to the parser. \code{add\_matter(name, n, reps, \dots)} declares an $n$-component multiplet, each gauge factor assigned a representation: \texttt{"fund"} (the defining representation), \texttt{"adj"}, a rational number (the charge under a $\mathrm U(1)$ factor) or an explicit matrix list. Spinors request a Dirac gamma-matrix representation. The parser's \verb|D_\mu| then acts on an operand by looking at its own representation data: on a multiplet it subtracts $\ii g\sum_{\text{groups}}A^a_\mu T^a_R$ for each factor, on an adjoint index it adds $gf^{abc}A^b_\mu(\cdot)^c$, and on a gauge-neutral field it reduces to $\pd_\mu$. Several factors compose, which is how $\mathrm{SU}(2)_L\times\mathrm U(1)_Y$ is built (\cref{ex:ew}).

\paragraph{Gauge variations and masses.} \code{gauge\_parameters(name)} returns arbitrary functions $\alpha^a(x)$; \code{gauge\_variation} turns them into a variation dictionary $\delta A^a_\mu=g^{-1}\pd_\mu\alpha^a+f^{abc}A^b_\mu\alpha^c$ and $\delta\phi=\ii\alpha^aT^a\phi$ for every matter multiplet, which can be passed directly to \code{symmetry\_defect} to test local invariance. \code{gauge\_mass\_matrix(L, vev)} returns the matrix $M^2_{ij}=\pd^2\mathcal L/\pd A_i^0\pd A_j^0$ at the vacuum \code{vev}, whose eigenvalues are the squared vector masses (the zero eigenvalues count unbroken generators) together with the basis labels.

\subsection{Expressing results covariantly: \code{fit}}
\label{sec:fit}

Component output such as $E^\nu$ for all four values of $\nu$ is correct but unreadable. \code{fit(model, target, basis)} receives a target tensor and a dictionary of named \LaTeX{} building blocks with the same free indices and returns coefficients $c_i$ such that $\text{target}=\sum_ic_i\,\text{block}_i$ componentwise, or \code{None} if no such combination exists. The implementation replaces each distinct derivative and function appearing in the components by a dummy symbol, treats the residual as a polynomial in these dummies, and solves for the $c_i$ the linear system obtained by equating all polynomial coefficients. Since \code{sympy.solve} silently ignores equations that do not involve the unknowns, the solution is \emph{re-substituted into every equation} and rejected if any residual remains. (An earlier version lacked this check and falsely accepted targets with extra terms; the negative-control test in the suite was added for this reason.) Examples of use appear throughout \cref{part:examples}: $\delta S/\delta A_\nu=\pd_\mu F^{\mu\nu}$ for Maxwell, $(\Box+\dots)$ for Podolsky, and the Belinfante relation.

\subsection{Grading}
\label{sec:grading}

The \code{teaching} module offers three graders returning a \code{Grade} object (truthy when correct, with a message).
\begin{itemize}[nosep]
\item \code{grade\_identity(model, text)} verifies a claimed identity \texttt{lhs = rhs} via \code{check}.
\item \code{grade\_field\_equation(model, L, submitted, field)} derives the correct equation of motion from $\mathcal L$ and uses \code{fit} to test whether the submission is a non-zero multiple of it. By default overall factors (including sign) are accepted, since equations of motion are defined up to normalisation; \code{allow\_overall\_factor=False} demands an exact match, and the message reports the factor when it is not $1$. Submitted strings may be written as \texttt{expr = 0}, \texttt{lhs = rhs}, or just \texttt{expr}.
\item \code{grade\_conserved\_current(model, L, variations, submitted, K)} computes the Noether current for the given symmetry and compares it with the submitted current.
\end{itemize}
Parse and index errors are caught and reported as ``could not evaluate'', so the module is robust against malformed submissions. It never passes student input to Python's \code{eval}.

\subsection{Testing and validation}
\label{sec:testing}

The repository contains $71$ automated tests: $53$ fast unit and integration tests organised by module (parser, tensors, variational calculus, Noether, Hamiltonian/waves, gauge, grading), and $18$ tests that run each example script and assert that no check printed \texttt{[FAIL]}. The unit tests include
\begin{itemize}[nosep]
\item parser tests: numbers and fractions, implicit multiplication, derivative scope, Einstein summation, $\Box$, $\epsilon$ and $\delta$, conjugation, strict-mode rejection of a repeated lower index, and error reporting;
\item tensor tests: raising and lowering in both signatures, symmetry predicates, Levi-Civita conventions, plain summation of spatial indices, and the covariant divergence of a vector on a curved background;
\item known results as oracles: the Klein--Gordon Hamiltonian, the Maxwell ($2$ dof) and Proca ($3$ dof, second class) canonical structure and dispersion, the electroweak masses, $T^{00}=\tfrac12(E^2+B^2)$ via the Hilbert tensor, the Belinfante relation, and the Dirac equation derived from the QED action;
\item negative controls: a non-symmetry is rejected by \code{noether\_current}, a wrong sign convention for the gauge field is detected as a non-invariance, the dilatation defect of a massive scalar is exactly $m^2\phi^2$, the grader rejects a wrong mass sign, and \code{fit} returns \code{None} for targets outside the span.
\end{itemize}
In addition, the examples exercise the library on problems with analytic answers that are not special-cased anywhere in the code: for instance the BPST action $8\pi^2/g^2$ and unit topological charge, or the monopole mass $4\pi v/g$ obtained by numerical quadrature of the energy density. Bugs found by this approach during development included mis-scoped derivatives, a spurious dispersion root, wrong sign conventions in the Higgs potential, and a vacuous symmetry test (above); each now has a regression test.

\subsection{Performance}
\label{sec:perf}

The cost is dominated by SymPy simplification of component expressions. On a single core of a cloud virtual machine, the fast unit suite takes about one minute and the full example suite about five minutes. Per-example wall-clock times are listed in \cref{tab:timings}; the sequential total is about $310$\,s. The $\mathrm{SU}(2)$ Yang--Mills Lagrangian expands to $180$ terms and its local gauge invariance is verified in about six seconds; the $\mathrm{SU}(3)$ Lagrangian has $1650$ terms. These figures illustrate the scaling; they are not a benchmark. The cost of component expansion grows as $d^{\,n_{\text{contracted}}}\dim(G)^{\,n_{\text{adj}}}$ for an expression with that many contracted spacetime and adjoint indices, which is why $\mathrm{SU}(N)$ with $N\ge4$ is slow.

\subsection{Availability}
\label{sec:avail}

The source is released under the MIT licence at \url{https://github.com/nikola-m/fieldsmith}. Installation is by \code{pip install -e .}; the only dependencies are SymPy and mpmath. The repository contains the library, the eighteen examples, the tests, the user guide, a teaching guide with a twelve-week syllabus and exercises, and the scripts that regenerate the example gallery and every number quoted in this paper (\cref{app:repro}).

\section{Worked examples}
\label{part:examples}

The eighteen examples below are the scripts in the \texttt{examples/} directory of the repository. Each follows the same pattern: the physical \emph{background} needed to understand the problem; the essential \emph{code} (the listings in this document are complete and were executed to produce the output quoted); the \emph{results} and their interpretation; and a suggested \emph{extension}. The full scripts contain further checks, each printed as \texttt{[PASS]} or \texttt{[FAIL]}; all examples pass in the released version, and their complete output is in the generated gallery. The examples are ordered roughly by sophistication; the grouping is thematic. Read in sequence, the backgrounds amount to a short introduction to classical field theory, from the scalar field to non-abelian monopoles.

\subsection{Scalar fields and solitons}

\exheading{ex:kg}{The Klein--Gordon field}{01\_klein\_gordon.py}{elementary}

\paragraph{Background.} The simplest relativistic field theory is a real scalar field $\phi(x)$ with
\begin{equation}
\mathcal L=\tfrac12\pd_\mu\phi\,\pd^\mu\phi-\tfrac12m^2\phi^2.
\label{eq:KG}
\end{equation}
The EL expression \cref{eq:EL} is $E=-(\Box\phi+m^2\phi)$, so the field equation is the Klein--Gordon equation $(\Box+m^2)\phi=0$. Plane waves $\phi\propto\ee^{-\ii(\omega t-\vect k\cdot\vect x)}$ solve it if $\omega^2=\vect k^2+m^2$, the relativistic mass shell; $m$ is the mass of the quanta in the quantum theory. Translation invariance gives the canonical tensor \cref{eq:Tcan},
\begin{equation}
T^\mu{}_\nu=\pd^\mu\phi\,\pd_\nu\phi-\delta^\mu_\nu\mathcal L,\qquad T^0{}_0=\tfrac12\big(\dot\phi^2+|\nabla\phi|^2+m^2\phi^2\big),
\end{equation}
whose $T^0{}_0$ is the energy density (positive-definite) and $T^i{}_0=-\dot\phi\,\pd_i\phi$ the energy flux. In the Hamiltonian description the momentum conjugate to $\phi$ is $\pi=\dot\phi$, the Hamiltonian density is $\mathcal H=\tfrac12\pi^2+\tfrac12|\nabla\phi|^2+\tfrac12m^2\phi^2$, and Hamilton's equations $\dot\phi=\delta H/\delta\pi=\pi$, $\dot\pi=-\delta H/\delta\phi=\nabla^2\phi-m^2\phi$ reproduce the second-order equation.

In the library:
\begin{lstlisting}
from fieldsmith import Model
from fieldsmith.waves import linearize, dispersion
from fieldsmith.noether import canonical_stress_tensor
from fieldsmith.hamiltonian import Hamiltonian

M = Model()                                   # 3+1 Minkowski, signature (+,-,-,-)
M.parameters("m", positive=True)
M.scalar("phi")
L = M.lagrangian(r"\frac12 \partial_\mu\phi\,\partial^\mu\phi - \frac12 m^2\phi^2")
E = M.euler_lagrange(L)                       # delta S / delta phi
print(E[0])                                   # -(Box phi + m^2 phi), componentwise
lin, pert = linearize(M, E)
print(dispersion(M, lin, list(pert.values()))["omega"])
T = canonical_stress_tensor(M, L)             # T^mu_nu
print(T[0, 0])                                # energy density
H = Hamiltonian(M, L)
print(H.H)
\end{lstlisting}

\paragraph{Results.} The EL expression comes out as $-m^2\phi-\pd_t^2\phi+\nabla^2\phi$, i.e.\ $-(\Box+m^2)\phi$. The linearisation and plane-wave substitution give $\det K=\omega^2-\vect k^2-m^2$ with roots $\omega=\pm\sqrt{\vect k^2+m^2}$. The canonical tensor component printed is $\tfrac12(m^2\phi^2+\dot\phi^2+|\nabla\phi|^2)$, and the Hamiltonian is written in terms of the momentum function \texttt{p[phi]}. The full example additionally asserts $E=-(\Box\phi+m^2\phi)$ via the parser and prints Hamilton's equations.

\paragraph{Extension.} Add a $\tfrac{\lambda}{4!}\phi^4$ term, repeat the analysis, and show that the dispersion relation of small oscillations about $\phi=0$ is unchanged while that about a non-zero background acquires a mass shift $\tfrac\lambda2\bar\phi^2$.

\exheading{ex:kink}{Solitons in $1{+}1$ dimensions: the $\phi^4$ kink and sine-Gordon}{02\_kink\_and\_sine\_gordon.py}{elementary}

\paragraph{Background.} In one space dimension, nonlinear scalar theories can have finite-energy, localised, static solutions called \emph{solitons} or \emph{kinks}. Consider
\begin{equation}
\mathcal L=\tfrac12(\pd\phi)^2-V(\phi),\qquad V=\tfrac\lambda4\Big(\phi^2-\tfrac{m^2}{\lambda}\Big)^2,
\end{equation}
which has two degenerate vacua $\phi=\pm v$, $v=m/\sqrt\lambda$. Finite energy requires the field to approach vacua at $x\to\pm\infty$; a solution interpolating between the two is topologically stable because it cannot be deformed to a vacuum without infinite cost, and carries the conserved topological charge $Q=[\phi(\infty)-\phi(-\infty)]/2v$. For static fields the energy is $E=\int\dd x\,[\tfrac12\phi'^2+V]$. Completing the square,
\begin{equation}
E=\int\dd x\,\tfrac12\big(\phi'\mp\sqrt{2V}\big)^2\pm\int\dd x\,\sqrt{2V}\,\phi'\ \ge\ \Big|\int_{-v}^{v}\sqrt{2V(\phi)}\,\dd\phi\Big|,
\label{eq:bogomolny1d}
\end{equation}
the \emph{Bogomolny bound} \cite{Bogomolny1976}, saturated by solutions of the first-order equation $\phi'=\pm\sqrt{2V}$. For the potential above, $\sqrt{2V}=\sqrt{\lambda/2}\,(v^2-\phi^2)$, the solution is the kink $\phi=v\tanh(mx/\sqrt2)$ and the bound gives the kink mass $M_K=2\sqrt2\,m^3/(3\lambda)$. A different potential, $V=1-\cos\phi$, gives the sine-Gordon equation $\phi_{tt}-\phi_{xx}+\sin\phi=0$, a completely integrable system with exact solutions: the (Lorentz-boosted) soliton $\phi=4\arctan\exp[\gamma(x-ut)]$ and the oscillating bound state, the \emph{breather}, $\phi=4\arctan\!\big[\tfrac{\sqrt{1-\omega^2}}{\omega}\sin(\omega t)/\cosh(\sqrt{1-\omega^2}\,x)\big]$ \cite{Rajaraman}.

In the library:
\begin{lstlisting}
import sympy as sp
from fieldsmith import Model, Spacetime

M = Model(Spacetime.minkowski(2))             # 1+1 dimensions
m, lam = M.parameters("m lambda", positive=True)
phi = M.scalar("phi")
x = M.coords[1]
V = r"\frac{\lambda}{4}\left(\phi^2 - \frac{m^2}{\lambda}\right)^2"
E = M.euler_lagrange(r"\frac12\partial_\mu\phi\,\partial^\mu\phi - " + V)
kink = (m / sp.sqrt(lam)) * sp.tanh(m * x / sp.sqrt(2))
print(E.check_solution({phi: kink}))          # True: exact kink
\end{lstlisting}

\paragraph{Results.} \code{check\_solution} returns \texttt{True}: the kink solves the full second-order EOM $-\lambda\phi^3+m^2\phi-\ddot\phi+\phi''=0$ exactly. The example also integrates $\int\sqrt{2V}\,\dd\phi$ symbolically to obtain $2\sqrt2m^3/3\lambda$, evaluates the energy density of the kink, $m^4/[2\lambda\cosh^4(mx/\sqrt2)]$ and integrates it over $x$, recovering the same number, which confirms the Bogomolny saturation. For sine-Gordon, with transcendental profiles, the residual is evaluated numerically at random points; the boosted soliton and the breather pass, and a soliton with the wrong velocity is rejected (negative control).

\paragraph{Extension.} Obtain the Bogomolny equation for a new potential (e.g.\ $V=\tfrac12(W')^2$ for a superpotential $W$) and verify that its solution satisfies the second-order equation.

\subsection{Abelian gauge fields}

\exheading{ex:maxwell}{Maxwell electrodynamics}{03\_maxwell.py}{intermediate}

\paragraph{Background.} The electromagnetic field is described by a four-potential $A_\mu$ with field strength $F_{\mu\nu}=\pd_\mu A_\nu-\pd_\nu A_\mu$ and
\begin{equation}
\mathcal L=-\tfrac14F_{\mu\nu}F^{\mu\nu}-J^\mu A_\mu,\qquad
-\tfrac14F_{\mu\nu}F^{\mu\nu}=\tfrac12(\vect E^2-\vect B^2),
\end{equation}
with $E_i=F_{i0}$ and $B_i=\tfrac12\epsilon_{ijk}F_{jk}$ in the conventions of the library. (Compared with the textbook fields $\vect E=-\nabla A^0-\pd_t\vect A$ and $\vect B=\nabla\times\vect A$, both carry an overall minus sign, which is immaterial for the quadratic combinations $\vect E^2\pm\vect B^2$ and $\vect E\!\cdot\!\vect B$ that occur below.) The EL expression is $\delta S/\delta A_\nu=\pd_\mu F^{\mu\nu}-J^\nu$, i.e.\ $\pd_\mu F^{\mu\nu}=J^\nu$ (Gauss's and Amp\`ere's laws). Two structural facts are essential. First, $F$ is automatically closed, $\pd_{[\mu}F_{\nu\rho]}=0$ (the Bianchi identity, i.e.\ the homogeneous Maxwell equations $\nabla\!\cdot\!\vect B=0$, $\nabla\times\vect E=-\pd_t\vect B$), equivalently $\pd_\mu\dual F^{\mu\nu}=0$ with $\dual F^{\mu\nu}=\tfrac12\epsilon^{\mu\nu\rho\sigma}F_{\rho\sigma}$. Second, the action is invariant under $A_\mu\to A_\mu+\pd_\mu\alpha$ with arbitrary $\alpha(x)$---a \emph{gauge} symmetry, whose Noether identity $\pd_\nu\pd_\mu F^{\mu\nu}\equiv0$ implies current conservation $\pd_\nu J^\nu=0$. The symmetric, gauge-invariant (Hilbert) energy--momentum tensor is $T^{\mu\nu}=-F^{\mu\alpha}F^\nu{}_\alpha+\tfrac14g^{\mu\nu}F^2$, with $T^{00}=\tfrac12(\vect E^2+\vect B^2)$, traceless in four dimensions (conformal invariance of Maxwell theory).

Canonically, $\pi^0=\pd\mathcal L/\pd\dot A_0=0$ is a primary constraint (no velocity of $A_0$ appears) and $\pi^i=-F^{0i}$ gives $\mathcal H=\tfrac12(\pi^2+\vect B^2)-A_0\,\nabla\!\cdot\!\vect\pi$; preservation of $\pi^0\weq0$ gives the secondary constraint $\nabla\!\cdot\!\vect\pi\weq0$ (Gauss's law). Both are first class and generate the gauge transformations, so with $N=4$, $N_1=2$, \cref{eq:dof} gives $n_{\mathrm{dof}}=2$: the two photon polarisations.

The listing from the introduction (\texttt{s03\_maxwell.py}) is reproduced here:
\begin{lstlisting}
from fieldsmith import Model
from fieldsmith.analysis import fit
from fieldsmith.hamiltonian import Hamiltonian

M = Model()
M.vector("A")
M.define(r"F_{\mu\nu}", r"\partial_\mu A_\nu - \partial_\nu A_\mu")
L = M.lagrangian(r"-\frac14 F_{\mu\nu}F^{\mu\nu}")
E = M.euler_lagrange(L)
print(fit(M, E.tensor("A"), {"dF": r"\partial_\mu F^{\mu\nu}"}))                  # {'dF': 1}
print(M.check(r"\partial_\nu\partial_\mu F^{\mu\nu} = 0"))                         # True
print(M.check(r"\partial_\mu (\frac12\epsilon^{\mu\nu\rho\sigma}F_{\rho\sigma}) = 0"))  # Bianchi
print(Hamiltonian(M, L).dirac_analysis().summary())
\end{lstlisting}
The remaining checks of the full example are: gauge invariance (\code{symmetry\_defect} for $\delta A_\mu=\pd_\mu\alpha$ returns $0$), the stress tensor ($T^{00}=\tfrac12(E^2+B^2)$ with $E,B$ built from components of $F$; traceless; symmetric), and the wave matrix.

\paragraph{Results.}
\begin{itemize}[nosep]
\item \code{fit} returns $\{\texttt{dF}: 1\}$: the EL expression is exactly $\pd_\mu F^{\mu\nu}$; with the source term $-J^\mu A_\mu$ it returns $\{\texttt{dF}:1,\ \texttt{J}:-1\}$.
\item Bianchi, its dual form and $\pd_\nu\pd_\mu F^{\mu\nu}=0$ are verified.
\item The Hilbert tensor satisfies $T^{00}=\tfrac12(\vect E^2+\vect B^2)$, is traceless and symmetric, while the canonical $T^0{}_0$ differs from it (they differ by a total derivative and a term vanishing on shell, \cref{ex:stress}).
\item The $4\times4$ plane-wave matrix has rank $3$ (not $4$): the null direction $a_\mu\propto k_\mu$ is pure gauge.
\item The Dirac analysis reports $4$ canonical pairs, $2$ first-class constraints, $0$ second-class and $2$ degrees of freedom; the constraints are $\pi^0\weq0$ (primary) and $\pd_i\pi^i\weq0$ (secondary).
\end{itemize}

\paragraph{Extension.} Add a Chern--Simons-like term or a mass term and watch the first-class constraints turn second class (\cref{ex:proca}).

\exheading{ex:proca}{Massive vector fields: Proca and St\"uckelberg}{04\_proca\_stueckelberg.py}{intermediate}

\paragraph{Background.} Giving the photon a mass breaks gauge invariance explicitly. The Proca Lagrangian \cite{Proca1936} is
\begin{equation}
\mathcal L=-\tfrac14F_{\mu\nu}F^{\mu\nu}+\tfrac12m^2A_\mu A^\mu,\qquad
\pd_\mu F^{\mu\nu}+m^2A^\nu=0.
\end{equation}
Taking the divergence, $m^2\pd_\nu A^\nu=0$: the Lorenz condition is a \emph{consequence} of the equation (not a gauge choice), and each component then satisfies $(\Box+m^2)A^\nu=0$, i.e.\ the mass shell with three independent polarisations (the transverse ones plus a longitudinal mode that Maxwell theory lacks). Canonically the mass term gives the secondary constraint $m^2A_0+\nabla\!\cdot\!\vect\pi\weq0$, which no longer commutes with $\pi^0\weq0$: both are second class, $N_2=2$ and $n_{\mathrm{dof}}=3$. Gauge invariance can be restored by introducing a scalar $\theta$ \cite{Stueckelberg1938},
\begin{equation}
\mathcal L=-\tfrac14F_{\mu\nu}F^{\mu\nu}+\tfrac12(mA_\mu-\pd_\mu\theta)(mA^\mu-\pd^\mu\theta),
\end{equation}
invariant under $\delta A_\mu=\pd_\mu\alpha$, $\delta\theta=m\alpha$. In the gauge $\theta=0$ (``unitary gauge'') the Proca theory is recovered, and the longitudinal polarisation is revealed as the scalar $\theta$, ``eaten'' by the photon. This is the abelian prototype of the Higgs mechanism (\cref{ex:higgs}).

In the library:
\begin{lstlisting}
from fieldsmith import Model
from fieldsmith.hamiltonian import Hamiltonian
M = Model()
M.parameters("m", positive=True)
M.vector("A")
M.define(r"F_{\mu\nu}", r"\partial_\mu A_\nu - \partial_\nu A_\mu")
L = M.lagrangian(r"-\frac14 F_{\mu\nu}F^{\mu\nu} + \frac12 m^2 A_\mu A^\mu")
print(Hamiltonian(M, L).dirac_analysis().dof)         # 3

S = Model()                                           # Stueckelberg form
S.parameters("m", positive=True)
S.vector("A"); S.scalar("theta")
S.define(r"F_{\mu\nu}", r"\partial_\mu A_\nu - \partial_\nu A_\mu")
Ls = S.lagrangian(r"-\frac14 F_{\mu\nu}F^{\mu\nu} + \frac12 (m A_\mu - \partial_\mu\theta)(m A^\mu - \partial^\mu\theta)")
print(Hamiltonian(S, Ls).dirac_analysis().summary())  # 2 first-class constraints, 3 dof
\end{lstlisting}

\paragraph{Results.} \code{fit} gives $\text{EL}=1\cdot\pd_\mu F^{\mu\nu}+1\cdot m^2A^\nu$, and the identity $\pd_\nu E^\nu\propto m^2\pd_\nu A^\nu$ is checked. The plane-wave determinant is $-m^2(\omega^2-\vect k^2-m^2)^3$ up to sign, so the mass-shell factor appears with exponent $3$: three polarisations. The Dirac analysis finds two second-class constraints and $3$ degrees of freedom. In the St\"uckelberg form, $\delta L=0$ under the combined transformation, there are $5$ canonical pairs, $2$ first-class constraints and $0$ second-class, and again $n_{\mathrm{dof}}=(10-4)/2=3$: the same physics in a gauge-invariant disguise.

\paragraph{Extension.} Couple the St\"uckelberg scalar to a conserved current and show that its longitudinal contribution to the exchange amplitude decouples as $m\to0$.

\exheading{ex:higgs}{Scalar QED and the abelian Higgs mechanism}{05\_abelian\_higgs.py}{intermediate}

\paragraph{Background.} A charged scalar $\Phi$ minimally coupled to the photon has the Lagrangian
\begin{equation}
\mathcal L=-\tfrac14F_{\mu\nu}F^{\mu\nu}+(D_\mu\Phi)^\dagger D^\mu\Phi+\mu^2\Phi^\dagger\Phi-\lambda(\Phi^\dagger\Phi)^2,\qquad D_\mu=\pd_\mu-\ii eA_\mu,
\end{equation}
invariant under the local $\mathrm U(1)$: $\Phi\to\ee^{\ii\alpha(x)}\Phi$, $A_\mu\to A_\mu+e^{-1}\pd_\mu\alpha$. The Noether current of the corresponding \emph{global} symmetry is the electromagnetic current, which appears on the right-hand side of Maxwell's equations: $\pd_\mu F^{\mu\nu}=-e\,j^\nu$, where $j^\mu=\ii\big(\Phi^\dagger D^\mu\Phi-(D^\mu\Phi)^\dagger\Phi\big)$. For $\mu^2>0$ the potential $V=-\mu^2|\Phi|^2+\lambda|\Phi|^4$ has a minimum away from zero, $|\Phi|^2=\mu^2/2\lambda$, and the symmetry is spontaneously broken: writing $\Phi=(v+h+\ii G)/\sqrt2$ with $v=\mu/\sqrt\lambda$, the real field $h$ has mass$^2$ $2\mu^2$, while $G$, the would-be Goldstone boson, mixes with the photon through $e\,v\,A^\mu\pd_\mu G$ (from the kinetic term). Choosing the unitary gauge $G=0$ removes $G$ and leaves a vector with mass $M_A=ev$ and a massive Higgs scalar \cite{Higgs1964,EnglertBrout1964,GHK1964}. Degrees of freedom are conserved: $2$ (photon) $+\,2$ (complex scalar) $\to 3$ (massive vector) $+\,1$ (Higgs).

In the library:
\begin{lstlisting}
from fieldsmith import Model
from fieldsmith.gauge import GaugeTheory, LieAlgebra
from fieldsmith.noether import symmetry_defect

M = Model()
G = GaugeTheory(M)
G.add_group("A", LieAlgebra.u1(), "e")                        # photon A_mu, coupling e
G.add_matter("Phi", 1, {"A": 1}, complex=True, latex=r"\Phi") # charge-1 complex scalar
L = M.lagrangian(r"-\frac14 F_{\mu\nu}F^{\mu\nu} + (D_\mu\Phi)^\dagger (D^\mu\Phi)"
                 r" + \mu^2\Phi^\dagger\Phi - \lambda(\Phi^\dagger\Phi)^2")
(alpha,) = G.gauge_parameters("A")
print(symmetry_defect(M, L, G.gauge_variation({"A": [alpha]})))   # 0: local U(1) invariance
\end{lstlisting}

\paragraph{Results.} Local invariance is verified ($\delta\mathcal L=0$). The Noether charge density of the global rotation is $J^0=-2eA_0\Phi\Phi^*+\ii(\Phi\dot\Phi^*-\Phi^*\dot\Phi)$ (in components) and the Noether identity holds. \code{fit} gives $\pd_\mu F^{\mu\nu}+e\,j^\nu$ for the photon equation. After substituting the vacuum, the coefficient of $A_x\pd_xG$ in $\mathcal L$ is $e(v+h)$ as expected. In the unitary gauge the plane-wave determinant factorises as $(\omega^2-\vect k^2-2\mu^2)\,[\omega^2-\vect k^2-e^2\mu^2/\lambda]^3$ (up to an overall constant), so the physical masses squared are $2\mu^2$ and $e^2\mu^2/\lambda=(ev)^2$ with multiplicities $1$ and $3$; the degree-of-freedom count is $4$ before and after.

\paragraph{Extension.} Obtain the vortex (Nielsen--Olesen) equations by assuming an axially symmetric ansatz and check the Bogomolny equations at $\lambda=e^2/2$.

\exheading{ex:goldstone}{The $\mathrm O(N)$ model and Goldstone's theorem}{06\_goldstone\_O\_N.py}{intermediate}

\paragraph{Background.} Goldstone's theorem \cite{Goldstone1961} states that for each broken generator of a continuous global symmetry there is a massless scalar. Take $N$ real scalars $\phi_p$ with
\begin{equation}
\mathcal L=\tfrac12\pd_\mu\phi_p\pd^\mu\phi_p+\tfrac{\mu^2}2\phi_p\phi_p-\tfrac\lambda4(\phi_p\phi_p)^2,
\end{equation}
invariant under $\mathrm O(N)$, with $N(N-1)/2$ generators. The vacuum manifold is the sphere $|\vec\phi|=v=\mu/\sqrt\lambda$; choosing a point, e.g.\ $\vec\phi=(0,\dots,0,v)$, breaks $\mathrm O(N)\to\mathrm O(N-1)$, so $\tfrac{N(N-1)}2-\tfrac{(N-1)(N-2)}2=N-1$ generators are broken. The mass matrix at the vacuum, $M^2_{pq}=\pd^2V/\pd\phi_p\pd\phi_q$, has one eigenvalue $2\mu^2$ (the radial ``Higgs'' mode) and $N-1$ zero eigenvalues (Goldstones). The Noether currents for the $N(N-1)/2$ rotations are $J^\mu_{pq}=\phi_p\pd^\mu\phi_q-\phi_q\pd^\mu\phi_p$.

\paragraph{In the library.} The index family $p,q,\dots$ is a user-defined index space; one Lagrangian string serves for any $N$.
\begin{lstlisting}
from fieldsmith import Model
N = 4
M = Model()
M.parameters("mu lambda", positive=True)
M.add_index_space("flavor", "pqrs", N)                  # the index family  p,q,r,s = 1..N
M.field("phi", ("flavor",), "")
L = M.lagrangian(r"\frac12\partial_\mu\phi_p\partial^\mu\phi_p + \frac{\mu^2}{2}\phi_p\phi_p"
                 r" - \frac{\lambda}{4}(\phi_p\phi_p)^2")
print(len(M.euler_lagrange(L)))                         # N field equations
\end{lstlisting}

\paragraph{Results.} For $N=2,3,4$ the example reports $1,3,6$ conserved currents; $1,2,3$ broken generators (those that move the vacuum); and mass spectra $\{0:1,\,2\mu^2:1\}$, $\{0:2,\,2\mu^2:1\}$, $\{0:3,\,2\mu^2:1\}$. Thus the number of massless Goldstone bosons equals $\dim\mathrm O(N)/\mathrm O(N-1)=N-1$ in each case (asserted).

\paragraph{Extension.} Add a small explicit symmetry-breaking term $-\epsilon\phi_N$ and compute the pseudo-Goldstone mass$^2$, $\epsilon/v$.

\subsection{Non-abelian gauge theory}

\exheading{ex:ym}{Yang--Mills theory: $\mathrm{SU}(2)$ and $\mathrm{SU}(3)$}{07\_yang\_mills.py}{advanced undergraduate / graduate}

\paragraph{Background.} Yang and Mills \cite{YangMills1954} generalised the abelian gauge principle to non-commuting symmetries. For a compact Lie group with generators $T^a$, $[T^a,T^b]=\ii f^{abc}T^c$, the gauge potential $A_\mu=A_\mu^aT^a$ transforms inhomogeneously and the field strength $F_{\mu\nu}=\pd_\mu A_\nu-\pd_\nu A_\mu-\ii g[A_\mu,A_\nu]$ has components given in \cref{eq:DF}. The non-linear term makes the gauge field \emph{self-interacting}: the Lagrangian $\mathcal L=-\tfrac14F^a_{\mu\nu}F^{a\mu\nu}$ contains cubic and quartic vertices ($\propto g$ and $\propto g^2$). The equation of motion is the covariant Maxwell equation
\begin{equation}
D_\mu F^{a\,\mu\nu}=\pd_\mu F^{a\,\mu\nu}+gf^{abc}A^b_\mu F^{c\,\mu\nu}=0,
\end{equation}
and the Bianchi identity $D_\mu\dual F^{a\,\mu\nu}=0$ holds identically. The curvature interpretation is encoded in
\begin{equation}
[D_\mu,D_\nu]\chi^a=g\,f^{abc}F^b_{\mu\nu}\chi^c
\label{eq:comm}
\end{equation}
for an adjoint field $\chi$. The group $\mathrm{SU}(N)$ has $N^2-1$ generators; in the defining representation $\tr(T^aT^b)=\tfrac12\delta^{ab}$ and the quadratic Casimir is $C_F=(N^2-1)/(2N)$, $3/4$ for $\mathrm{SU}(2)$ and $4/3$ for $\mathrm{SU}(3)$ (the QCD colour factor). A new feature compared with electromagnetism is that the field itself carries charge: the Noether current of the \emph{global} colour symmetry contains the gluon field and is conserved on shell.

In the library:
\begin{lstlisting}
from fieldsmith import Model
from fieldsmith.gauge import GaugeTheory, LieAlgebra

su3 = LieAlgebra.su(3)
print(su3.dim, su3.jacobi_holds(), su3.casimir_fundamental()[0, 0])   # 8 True 4/3
M = Model()
G = GaugeTheory(M)
G.add_group("A", LieAlgebra.su(2), "g")
L = M.lagrangian(r"-\frac14 F^a_{\mu\nu}F^{a\mu\nu}")
print(M.check(r"D_\mu(\frac12\epsilon^{\mu\nu\rho\sigma}F^a_{\rho\sigma}) = 0"))   # Bianchi, non-abelian
\end{lstlisting}

\paragraph{Results.} The library constructs $\mathrm{SU}(2)$ and $\mathrm{SU}(3)$ from their Gell-Mann matrices; the Jacobi identity and the Casimirs $C_F=3/4$ and $4/3$ are verified; the nonzero $f^{abc}$ of $\mathrm{SU}(3)$ come out as $f^{123}=1$, $f^{147}=f^{246}=f^{257}=f^{345}=\tfrac12$, $f^{156}=f^{367}=-\tfrac12$, $f^{458}=f^{678}=\tfrac{\sqrt3}2$, the standard table. The $\mathrm{SU}(2)$ Lagrangian has $180$ terms (kinetic, cubic, quartic) with highest power $g^2$. Local gauge invariance under $\delta A^a_\mu=g^{-1}\pd_\mu\alpha^a+f^{abc}A^b_\mu\alpha^c$ is verified: $\delta\mathcal L=0$ (about six seconds). The EL expression is $D_\mu F^{a\mu\nu}$ with coefficient $1$ (via \code{fit}), the Bianchi identity and the off-shell identity $D_\nu D_\mu F^{a\mu\nu}=0$ are confirmed, and the commutator relation \cref{eq:comm} holds for an adjoint multiplet $\chi$. For $\mathrm{SU}(3)$ the example builds the $8\times4=32$ gauge field components, a Lagrangian with $1650$ terms, and verifies the Bianchi identity.

\paragraph{Extension.} Linearise about the vacuum to show that the gauge bosons are massless with two polarisations, then find a non-trivial constant solution of $D_\mu F^{\mu\nu}=0$ (a ``Wu--Yang'' configuration).

\exheading{ex:ew}{The electroweak Higgs mechanism, $\mathrm{SU}(2)_L\times\mathrm U(1)_Y$}{08\_electroweak.py}{graduate}

\paragraph{Background.} The electroweak theory \cite{Glashow1961,Weinberg1967,Salam1968} has gauge group $\mathrm{SU}(2)_L\times\mathrm U(1)_Y$ with fields $W^a_\mu$ (coupling $g$, field strength $V^a_{\mu\nu}$) and $B_\mu$ (coupling $g'$, field strength $H_{\mu\nu}$), and a Higgs doublet $\Phi$ of hypercharge $Y=\tfrac12$:
\begin{equation}
\mathcal L=-\tfrac14V^a_{\mu\nu}V^{a\mu\nu}-\tfrac14H_{\mu\nu}H^{\mu\nu}+(D_\mu\Phi)^\dagger D^\mu\Phi+\mu^2\Phi^\dagger\Phi-\lambda(\Phi^\dagger\Phi)^2,\quad
D_\mu=\pd_\mu-\ii gW^a_\mu\tfrac{\sigma^a}{2}-\ii g'\tfrac12B_\mu.
\end{equation}
Choosing the vacuum $\langle\Phi\rangle=(0,v/\sqrt2)^T$, $v=\mu/\sqrt\lambda$, the kinetic term yields a vector-boson mass term $\tfrac12M^2_{ij}A^i_\mu A^{j\mu}$ in the basis $(W^1,W^2,W^3,B)$ with
\begin{equation}
M^2=\frac{v^2}4\begin{pmatrix}g^2&0&0&0\\0&g^2&0&0\\0&0&g^2&-gg'\\0&0&-gg'&g'^2\end{pmatrix}.
\end{equation}
The eigenvalues are $M_W^2=g^2v^2/4$ (twice: $W^\pm=(W^1\mp\ii W^2)/\sqrt2$), $M_Z^2=(g^2+g'^2)v^2/4$ and $0$. The massless eigenvector is the photon, $A=\sin\theta_W\,W^3+\cos\theta_W\,B$ with $\tan\theta_W=g'/g$, orthogonal to $Z=\cos\theta_W\,W^3-\sin\theta_W\,B$; the electric charge is $e=g\sin\theta_W=g'\cos\theta_W$. The tree-level relation $\rho\equiv M_W^2/(M_Z^2\cos^2\theta_W)=1$ is a consequence of the custodial $\mathrm{SU}(2)$ symmetry of the Higgs potential. Three of the four real Higgs components are eaten by $W^\pm,Z$; the fourth is the physical Higgs boson.

\paragraph{In the library.} Two gauge factors are added, and the Higgs doublet transforms under both: in the fundamental of $\mathrm{SU}(2)$ and with hypercharge $\tfrac12$.
\begin{lstlisting}
import sympy as sp
from fieldsmith import Model
from fieldsmith.gauge import GaugeTheory, LieAlgebra

M = Model()
G = GaugeTheory(M)
G.add_group("W", LieAlgebra.su(2), "g", strength="V")
G.add_group("B", LieAlgebra.u1(), "g'", strength="H")
G.add_matter("Phi", 2, {"W": "fund", "B": sp.Rational(1, 2)}, complex=True, latex=r"\Phi")
L = M.lagrangian(r"-\frac14 V^a_{\mu\nu}V^{a\mu\nu} - \frac14 H_{\mu\nu}H^{\mu\nu}"
                 r" + (D_\mu\Phi)^\dagger (D^\mu\Phi) + \mu^2\Phi^\dagger\Phi - \lambda(\Phi^\dagger\Phi)^2")
v = M.parameters("v", positive=True)
Phi = M.tensors["Phi"].data[()]
vev = {}
for f, val in zip(list(Phi), (0, v / sp.sqrt(2))):          # <Phi> = (0, v/sqrt2)
    vev[f] = val
    vev[M._conj_subs[f]] = val
M2, labels = G.gauge_mass_matrix(M.lagrangian(r"(D_\mu\Phi)^\dagger (D^\mu\Phi)"), vev)
print(labels, M2.eigenvals())        # {g^2 v^2/4: 2, (g^2+g'^2) v^2/4: 1, 0: 1}
\end{lstlisting}

\paragraph{Results.} The full Lagrangian is invariant under the four independent local parameters ($\delta\mathcal L=0$). The mass matrix is exactly the one above, with eigenvalues $\{g^2v^2/4:2,\ (g^2+g'^2)v^2/4:1,\ 0:1\}$. The example asserts two $W$ bosons with $M_W^2=g^2v^2/4$, one $Z$ with $M_Z^2=(g^2+g'^2)v^2/4$ and a massless photon; it extracts $\cos\theta_W=M_W/M_Z=g/\sqrt{g^2+g'^2}$, finds the photon direction as the null vector $(0,0,g'/g,1)$, i.e.\ $A\propto g'W^3+gB$, and verifies $\rho=1$.

\paragraph{Extension.} Add a second Higgs doublet (two-Higgs-doublet model) and compute $\rho$ for different vacuum alignments; add a Higgs triplet with $Y=1$ and show that $\rho\ne1$ at tree level.

\subsection{Matter, energy--momentum and gravity}

\exheading{ex:qed}{Quantum electrodynamics with a classical Dirac field}{09\_qed\_dirac.py}{graduate}

\paragraph{Background.} The Dirac field is a four-component spinor $\psi$ obeying $(\ii\gamma^\mu\pd_\mu-m)\psi=0$, with gamma matrices satisfying $\{\gamma^\mu,\gamma^\nu\}=2g^{\mu\nu}$. Its Lagrangian is $\bar\psi(\ii\gamma^\mu\pd_\mu-m)\psi$ with $\bar\psi=\psi^\dagger\gamma^0$. Coupling to electromagnetism by minimal substitution $\pd_\mu\to D_\mu=\pd_\mu-\ii eA_\mu$ gives
\begin{equation}
\mathcal L_{\mathrm{QED}}=\bar\psi(\ii\gamma^\mu D_\mu-m)\psi-\tfrac14F_{\mu\nu}F^{\mu\nu}.
\end{equation}
Variation with respect to $\bar\psi$ gives the Dirac equation, and with respect to $A_\nu$ gives $\pd_\mu F^{\mu\nu}=-e\,\bar\psi\gamma^\nu\psi$. The phase rotation $\psi\to\ee^{\ii\alpha}\psi$ yields the Noether current $J^\mu=-\bar\psi\gamma^\mu\psi$ in the sign convention of the library, which is the electromagnetic current (up to sign and the factor $e$). Applying $(\ii\gamma\cdot\pd+m)$ to the Dirac equation gives $-(\Box+m^2)\psi=0$: each component is a Klein--Gordon field. Canonically, the Lagrangian is linear in the velocities, so every momentum is a constraint: $\pi_\psi=\ii\psi^\dagger$ and $\pi_{\psi^\dagger}=0$, all second class. With eight real-equivalent fields $(\psi,\psi^*)$ this leaves $(16-8)/2=4$ degrees of freedom; in the quantum theory these are the two spin states of the particle and of the antiparticle. We treat $\psi$ here as a \emph{classical commuting} field; Fermi statistics enters only in quantisation.

In the library:
\begin{lstlisting}
from fieldsmith import Model
from fieldsmith.gauge import GaugeTheory, LieAlgebra
M = Model()
G = GaugeTheory(M)
G.add_group("A", LieAlgebra.u1(), "e")
G.add_matter("psi", 4, {"A": 1}, complex=True, spinor=True, latex=r"\psi")
M.parameters("m", positive=True)
L = M.lagrangian(r"\bar\psi (i\gamma^\mu D_\mu - m)\psi - \frac14 F_{\mu\nu}F^{\mu\nu}")
E = M.euler_lagrange(L)
from fieldsmith.analysis import fit
from fieldsmith.noether import noether_current, complex_variation
print(fit(M, E.tensor("A"), {"dF": r"\partial_\mu F^{\mu\nu}", "j": r"e \bar\psi\gamma^\nu\psi"}))
nc = noether_current(M, L, complex_variation(M, "psi", r"i\psi"))   # global phase rotation
print(nc.identity_holds())
\end{lstlisting}

\paragraph{Results.} Gauge invariance under $\psi\to\ee^{\ii\alpha}\psi$, $A\to A+\pd\alpha/e$ holds ($\delta\mathcal L=0$). The variation with respect to $\psi^\dagger$ equals $\gamma^0(\ii\gamma^\mu D_\mu-m)\psi$ (asserted). \code{fit} gives $\pd_\mu F^{\mu\nu}+e\bar\psi\gamma^\nu\psi$ with unit coefficients. The Noether current equals $-\bar\psi\gamma^\mu\psi$, and the Noether identity holds. The squared Dirac operator gives $-(\Box+m^2)\psi$ (asserted). The canonical analysis of the free Dirac Lagrangian finds $8$ second-class constraints and $4$ degrees of freedom.

\paragraph{Extension.} Add a Pauli term $\tfrac{\kappa}{2}\bar\psi\sigma^{\mu\nu}\psi F_{\mu\nu}$ (anomalous magnetic moment) and read off its contribution to the current.

\exheading{ex:stress}{Energy--momentum tensors: canonical, Belinfante and Hilbert}{10\_stress\_tensors.py}{graduate}

\paragraph{Background.} Translation invariance gives the canonical tensor \cref{eq:Tcan}, which for gauge fields is neither symmetric nor gauge invariant. For the Maxwell field, $T^{\mu\nu}_{\mathrm{can}}=-F^{\mu\alpha}\pd^\nu A_\alpha+\tfrac14g^{\mu\nu}F^2$. The Hilbert tensor \cref{eq:Thilb} is $T^{\mu\nu}_{\mathrm H}=-F^{\mu\alpha}F^\nu{}_\alpha+\tfrac14g^{\mu\nu}F^2$. Their difference is
\begin{equation}
T_{\mathrm H}^{\mu\nu}-T_{\mathrm{can}}^{\mu\nu}=F^{\mu\lambda}\pd_\lambda A^\nu=\pd_\lambda\big(F^{\mu\lambda}A^\nu\big)-\big(\pd_\lambda F^{\mu\lambda}\big)A^\nu,
\label{eq:belinfante}
\end{equation}
which is the Belinfante construction \cite{Belinfante1940,Rosenfeld1940}: the second term vanishes on shell (vacuum Maxwell equation) and the first is a total derivative of a quantity antisymmetric in $(\mu,\lambda)$, which does not change the conserved charges $P^\nu$ but makes the tensor gauge invariant and symmetric. A second ambiguity concerns the massless scalar: conformal invariance in $d=4$ is not manifest in $T_{\mathrm{can}}$, whose trace $T^\mu{}_\mu=-(\pd\phi)^2$ does not vanish. The ``improved'' tensor \cite{CallanColemanJackiw1970}
\begin{equation}
T^{\mu\nu}_{\mathrm{imp}}=T^{\mu\nu}-\tfrac16\big(\pd^\mu\pd^\nu-g^{\mu\nu}\Box\big)\phi^2
\end{equation}
is still conserved and has the same charges but trace $T^\mu{}_\mu\big|_{\mathrm{imp}}=\phi\Box\phi$, which vanishes on shell.

In the library:
\begin{lstlisting}
from fieldsmith import Model
from fieldsmith.analysis import fit
from fieldsmith.noether import canonical_stress_tensor, hilbert_stress_tensor
M = Model()
M.vector("A")
M.define(r"F_{\mu\nu}", r"\partial_\mu A_\nu - \partial_\nu A_\mu")
LM = r"-\frac14 F_{\mu\nu}F^{\mu\nu}"
Tc = canonical_stress_tensor(M, LM).raise_("nu")
Th = hilbert_stress_tensor(M, LM).raise_("mu").raise_("nu")
print(Tc.is_symmetric("mu", "nu"), Th.is_symmetric("mu", "nu"))     # False True
print(fit(M, Th - Tc.relabel(["mu", "nu"]),
          {"b": r"\partial_\lambda(F^{\mu\lambda}A^\nu)", "e": r"(\partial_\lambda F^{\mu\lambda})A^\nu"},
          labels=["mu", "nu"]))
\end{lstlisting}

\paragraph{Results.} The canonical tensor is not symmetric (\texttt{False}), the Hilbert tensor is (\texttt{True}); and \code{fit} returns $\{\texttt{b}:1,\ \texttt{e}:-1\}$, which is \cref{eq:belinfante}. For the scalar, the example prints the two traces: $-\dot\phi^2+|\nabla\phi|^2$ for the ordinary tensor and $\phi\,\Box\phi$ (in components, $\phi(\ddot\phi-\nabla^2\phi)$) for the improved tensor.

\paragraph{Extension.} Compute $T_{\mathrm H}^{\mu\nu}$ of the Proca field and show that the trace is $-m^2A^2$, i.e.\ conformal invariance is broken exactly by the mass.

\exheading{ex:frw}{Fields on curved backgrounds: FRW cosmology and Schwarzschild}{11\_curved\_spacetime.py}{graduate}

\paragraph{Background.} In a curved spacetime the action carries $\sqrt{|g|}$ and derivatives become covariant. For a spatially flat Friedmann--Robertson--Walker universe, $\dd s^2=\dd t^2-a(t)^2\dd\vect x^2$, with Hubble rate $H=\dot a/a$, the Einstein tensor has $G_{00}=3H^2$, and a minimally coupled scalar satisfies
\begin{equation}
\ddot\phi+3H\dot\phi-\frac{\nabla^2\phi}{a^2}+m^2\phi=0,
\end{equation}
the Klein--Gordon equation with a friction term due to expansion. The energy--momentum tensor $T_{\mu\nu}=\pd_\mu\phi\pd_\nu\phi-g_{\mu\nu}\mathcal L$ is covariantly conserved on shell, $\nabla_\mu T^{\mu\nu}=(\Box\phi+m^2\phi)\pd^\nu\phi$ (with $\Box=\nabla^\mu\nabla_\mu$). For a homogeneous field, the $00$ Einstein equation $G_{00}=8\pi GT_{00}$ is the Friedmann equation $3H^2=8\pi G(\tfrac12\dot\phi^2+\tfrac12m^2\phi^2)$. The Schwarzschild metric is a vacuum solution, $G_{\mu\nu}=0$.

\paragraph{In the library.} The metric is entered as a SymPy matrix; $G_{\mu\nu}$ is computed from the Christoffel symbols and the Riemann tensor.
\begin{lstlisting}
import sympy as sp
from fieldsmith import Model, Spacetime
t, x, y, z = sp.symbols("t x y z", real=True)
a = sp.Function("a", positive=True)(t)
frw = Spacetime([t, x, y, z], sp.diag(1, -a**2, -a**2, -a**2))
print(sp.simplify(frw.einstein()[0, 0]))                 # 3 a'^2 / a^2
M = Model(frw)
M.parameters("m", positive=True)
M.scalar("phi")
E = M.euler_lagrange(r"\frac12\partial_\mu\phi\,\partial^\mu\phi - \frac12 m^2\phi^2")
M.define(r"T_{\mu\nu}", r"\partial_\mu\phi\,\partial_\nu\phi - g_{\mu\nu}"
         r"\left(\frac12\partial_\alpha\phi\,\partial^\alpha\phi - \frac12 m^2\phi^2\right)")
print(M.check(r"\nabla_\mu T^{\mu\nu} = (\Box\phi + m^2\phi)\,\partial^\nu\phi"))
\end{lstlisting}

\paragraph{Results.} $G_{00}=3\dot a^2/a^2$ (asserted); the equation of motion equals the one above (asserted); \verb|\nabla_\mu T^{\mu\nu} = (\Box\phi + m^2\phi)\partial^\nu\phi| is verified by \code{check}; the homogeneous energy density is $\tfrac12\dot\phi_h^2+\tfrac12m^2\phi_h^2$, so $G_{00}=8\pi GT_{00}$ is the Friedmann equation. The example also confirms that the Schwarzschild Einstein tensor and Ricci scalar vanish.

\paragraph{Extension.} Derive the equation for tensor perturbations of a scalar-dominated universe, or add a non-minimal coupling $-\tfrac12\xi R\phi^2$ and find the conformal value $\xi=1/6$ at which the massless field is conformally invariant.

\subsection{Towards the research frontier}
\label{sec:frontier}

The remaining examples show how the same machinery handles theories that are not in standard textbooks or that probe current research topics. In each case the point is that a modification of a known Lagrangian can be explored in a few lines.

\exheading{ex:axion}{Axion electrodynamics and topological terms}{12\_axion\_electrodynamics.py}{research-flavoured}

\paragraph{Background.} Pseudoscalar (``axion-like'') fields $a$ couple to photons through the topological density $F_{\mu\nu}\dual F^{\mu\nu}$ \cite{Wilczek1987,CarrollFieldJackiw1990}:
\begin{equation}
\mathcal L=-\tfrac14F_{\mu\nu}F^{\mu\nu}+\tfrac12\pd_\mu a\,\pd^\mu a-\tfrac12m_a^2a^2-\tfrac g4\,a\,F_{\mu\nu}\dual F^{\mu\nu}.
\end{equation}
With $\epsilon^{0123}=+1$ one has $F_{\mu\nu}\dual F^{\mu\nu}=-4\,\vect E\!\cdot\!\vect B$, so the interaction is $g\,a\,\vect E\!\cdot\!\vect B$. The density is a total derivative, $F\dual F=\pd_\mu K^\mu$ with the Chern--Simons current $K^\mu=2\epsilon^{\mu\nu\rho\sigma}A_\nu\pd_\rho A_\sigma$; consequently, for a \emph{constant} $a$ the coupling has no effect on the equations of motion. The EL equations are
\begin{equation}
\pd_\mu F^{\mu\nu}+g\,\pd_\mu a\,\dual F^{\mu\nu}=0,\qquad(\Box+m_a^2)\,a=-\tfrac g4F_{\mu\nu}\dual F^{\mu\nu}=g\,\vect E\!\cdot\!\vect B,
\end{equation}
where we used $\pd_\mu\dual F^{\mu\nu}=0$. For a slowly rolling background $\pd_t a=\kappa$ the photon equation becomes $\pd_\mu F^{\mu\nu}+g\kappa\,\dual F^{0\nu}=0$, which is parity-violating: the two circular polarisations obey different dispersion relations,
\begin{equation}
\omega^2=k^2\pm g\kappa k,
\end{equation}
so that the plane of linear polarisation rotates by $\Delta\theta=\tfrac12g\,\Delta a$ after propagation---\emph{cosmic birefringence}, a signature searched for in the cosmic microwave background.

In the library:
\begin{lstlisting}
from fieldsmith import Model
from fieldsmith.analysis import fit
M = Model()
M.parameters("m_a g", positive=True)
M.vector("A"); M.scalar("a")
M.define(r"F_{\mu\nu}", r"\partial_\mu A_\nu - \partial_\nu A_\mu")
M.define(r"\tilde{F}^{\mu\nu}", r"\frac12\epsilon^{\mu\nu\rho\sigma}F_{\rho\sigma}")
L = M.lagrangian(r"-\frac14 F_{\mu\nu}F^{\mu\nu} + \frac12\partial_\mu a\partial^\mu a"
                 r" - \frac12 m_a^2 a^2 - \frac{g}{4} a F_{\mu\nu}\tilde F^{\mu\nu}")
E = M.euler_lagrange(L)
print(fit(M, E.tensor("A"), {"dF": r"\partial_\mu F^{\mu\nu}", "ax": r"\partial_\mu a\,\tilde F^{\mu\nu}"}))
print(M.check(r"F_{\mu\nu}\tilde F^{\mu\nu} = \partial_\mu(2\epsilon^{\mu\nu\rho\sigma}A_\nu\partial_\rho A_\sigma)"))
\end{lstlisting}

\paragraph{Results.} \code{fit} returns $\{\texttt{dF}:1,\ \texttt{ax}:g\}$ for the photon equation, and the identity $F\dual F=\pd_\mu(2\epsilon^{\mu\nu\rho\sigma}A_\nu\pd_\rho A_\sigma)$ is verified; the example also checks $F\dual F=-4\vect E\!\cdot\!\vect B$ and the axion equation. The $2\times2$ transverse plane-wave matrix for a wave along $z$ in the background $a=\kappa t$ is $\begin{pmatrix}\omega^2-k_z^2&-\ii g\kappa k_z\\ \ii g\kappa k_z&\omega^2-k_z^2\end{pmatrix}$ with determinant $-(g\kappa k_z-k_z^2+\omega^2)(g\kappa k_z+k_z^2-\omega^2)$, hence $\omega^2=k_z^2\pm g\kappa k_z$.

\paragraph{Extension.} Add a background magnetic field and compute the photon--axion mixing matrix, the basis of laboratory axion searches.

\exheading{ex:eh}{Nonlinear electrodynamics: vacuum birefringence}{13\_vacuum\_birefringence.py}{research-flavoured}

\paragraph{Background.} Quantum corrections from virtual electron--positron pairs make the vacuum nonlinear. At energies far below $m_e$ they are summarised by the Euler--Heisenberg effective Lagrangian \cite{EulerHeisenberg1936}; its weak-field expansion is
\begin{equation}
\mathcal L=\tfrac12(\vect E^2-\vect B^2)+\frac{2\alpha^2}{45m_e^4}\Big[(\vect E^2-\vect B^2)^2+7(\vect E\!\cdot\!\vect B)^2\Big].
\end{equation}
Using $F^2=-2(\vect E^2-\vect B^2)$ and $F\dual F=-4\vect E\!\cdot\!\vect B$ this is of the form $\mathcal L=-\tfrac14F^2+c_1(F^2)^2+c_2(F\dual F)^2$ with $c_1=\tfrac{\alpha^2}{90m_e^4}$ and $c_2=\tfrac{7\alpha^2}{360m_e^4}$. Light propagating through a uniform magnetic field $\vect B_0$ is therefore birefringent: the two linear polarisations have different refractive indices \cite{BialynickaBirula1970,Adler1971}. In general, linearising about $\vect B_0$ and propagating perpendicular to it, a photon with $\vect E\perp\vect B_0$ and one with $\vect E\parallel\vect B_0$ have
\begin{equation}
n_\perp-1=16\,c_1B_0^2,\qquad n_\parallel-1=16\,c_2B_0^2.
\end{equation}
With the Euler--Heisenberg coefficients, $n_\perp-1=\frac{8\alpha^2B_0^2}{45m_e^4}$ and $n_\parallel-1=\frac{14\alpha^2B_0^2}{45m_e^4}$, a ratio of $7/4$ that is independent of $B_0$. In terms of the critical field $B_{\mathrm{cr}}=m_e^2/e$ this is $n-1=\frac{\alpha}{4\pi}(B_0/B_{\mathrm{cr}})^2\times\{\frac{8}{45},\frac{14}{45}\}$.

\paragraph{In the library.} The EL equations of a cubic-in-field-strength theory (348 terms in $E^0$) are linearised about the background $A_y=B_0x$.
\begin{lstlisting}
from fieldsmith import Model
M = Model()
M.parameters("c_1 c_2", positive=True)
M.vector("A")
M.define(r"F_{\mu\nu}", r"\partial_\mu A_\nu - \partial_\nu A_\mu")
M.define(r"\tilde{F}^{\mu\nu}", r"\frac12\epsilon^{\mu\nu\rho\sigma}F_{\rho\sigma}")
L = M.lagrangian(r"-\frac14 F_{\mu\nu}F^{\mu\nu} + c_1 (F_{\mu\nu}F^{\mu\nu})^2 + c_2 (F_{\mu\nu}\tilde F^{\mu\nu})^2")
E = M.euler_lagrange(L)
import sympy as sp
from fieldsmith.waves import linearize, dispersion
B0 = M.parameters("B_0", positive=True)
x = M.coords[1]
A = M.tensors["A"].funcs
lin, pert = linearize(M, E, background={A[(2,)]: B0 * x})     # uniform B_0 along z
res = dispersion(M, lin, list(pert.values()))
w, ky = res["wave"].omega, res["wave"].k[1]
K = res["K"].subs({res["wave"].k[0]: 0, res["wave"].k[2]: 0})  # propagate along y
print(sp.factor(K[1, 1]), sp.factor(K[3, 3]))                   # the two linear polarisations
\end{lstlisting}

\paragraph{Results.} For propagation along $y$ the two linear polarisations decouple. For $A_x$ ($\vect E\perp\vect B_0$) the dispersion equation is $(1-16c_1B_0^2)\,\omega^2-(1-48c_1B_0^2)\,k_y^2=0$, giving $n=k_y/\omega=1+16c_1B_0^2+\order{B_0^4}$; for $A_z$ ($\vect E\parallel\vect B_0$) it is $(1-16c_1B_0^2+32c_2B_0^2)\,\omega^2-(1-16c_1B_0^2)\,k_y^2=0$, giving $n=1+16c_2B_0^2+\order{B_0^4}$. Hence $n_\perp-1=16c_1B_0^2$ and $n_\parallel-1=16c_2B_0^2$. Inserting the Euler--Heisenberg coefficients, the library prints $n_\perp-1=8\alpha^2B_0^2/(45m_e^4)$, $n_\parallel-1=14\alpha^2B_0^2/(45m_e^4)$ and the ratio $7/4$ (asserted). The labelling of which polarisation is ``parallel'' is a convention, and the example asserts the ratio is $7/4$ or $4/7$.

\paragraph{Extension.} Born--Infeld electrodynamics, expanded to quartic order, has $c_1=c_2=1/(32b^2)$ (with $b$ the maximal field strength). Setting $c_1=c_2$ in the result above gives $n_\perp=n_\parallel$ at this order: Born--Infeld theory is not birefringent, a property that singles it out among nonlinear electrodynamics. Verify this and explore deformations.

\exheading{ex:podolsky}{Higher-derivative electrodynamics: Podolsky's theory}{14\_podolsky\_higher\_derivative.py}{research-flavoured}

\paragraph{Background.} The self-energy of a point charge in Maxwell theory diverges. Podolsky \cite{Podolsky1942} proposed adding a term with second derivatives of $A_\mu$,
\begin{equation}
\mathcal L=-\tfrac14F_{\mu\nu}F^{\mu\nu}+\tfrac{a^2}2\,\pd^\mu F_{\mu\nu}\,\pd_\lambda F^{\lambda\nu},
\end{equation}
which remains gauge invariant and Lorentz invariant but has \emph{fourth-order} field equations,
\begin{equation}
(1+a^2\Box)\,\pd_\mu F^{\mu\nu}=J^\nu.
\end{equation}
Operators factorise: besides the massless photon there are massive modes with mass $1/a$. The linearised wave matrix has determinant (after removing gauge directions) $(\omega^2-\vect k^2)^2\,[\,1+a^2(\vect k^2-\omega^2)\,]^3$: two massless polarisations and three of mass $1/a$, so the theory propagates $5$ degrees of freedom \cite{HenneauxTeitelboim}. The extra modes are a hallmark of higher-derivative theories (Ostrogradsky's instability \cite{Ostrogradsky1850,Woodard2007}): they enter with a sign that is wrong for the energy, or for the norm of states in the quantum theory, so the theory is usually considered an effective one for $E\ll1/a$. The attractive feature is the Coulomb potential. For a static point charge, $(1-a^2\nabla^2)\nabla^2A_0=-q\,\delta^3(\vect x)$ (with $\Box\to-\nabla^2$), and partial fractions $\tfrac1{k^2(1+a^2k^2)}=\tfrac1{k^2}-\tfrac{a^2}{1+a^2k^2}$ give
\begin{equation}
A_0(r)=\frac{q}{4\pi r}\big(1-\ee^{-r/a}\big),
\end{equation}
which reduces to Coulomb's law for $r\gg a$ and is \emph{finite} at the origin, $A_0(0)=q/(4\pi a)$.

\paragraph{In the library.} The tensor $j_\nu=\pd^\mu F_{\mu\nu}$ is defined first (it contains second derivatives of the potential), after which the Lagrangian is a one-liner. \code{euler\_lagrange} handles the fourth-order equations automatically.
\begin{lstlisting}
from fieldsmith import Model
from fieldsmith.analysis import fit
from fieldsmith.waves import linearize, dispersion
M = Model()
M.parameters("a", positive=True)
M.vector("A")
M.define(r"F_{\mu\nu}", r"\partial_\mu A_\nu - \partial_\nu A_\mu")
M.define(r"j_\nu", r"\partial^\mu F_{\mu\nu}")
L = M.lagrangian(r"-\frac14 F_{\mu\nu}F^{\mu\nu} + \frac{a^2}{2} j_\nu j^\nu")
E = M.euler_lagrange(L)                                   # fourth order
print(fit(M, E.tensor("A"), {"dF": r"\partial_\mu F^{\mu\nu}", "box": r"\Box\partial_\mu F^{\mu\nu}"}))
lin, pert = linearize(M, E)
print(dispersion(M, lin, list(pert.values()))["pdet"])
\end{lstlisting}

\paragraph{Results.} \code{fit} returns $\{\texttt{dF}:1,\ \texttt{box}:a^2\}$, i.e.\ the field equation is $(1+a^2\Box)\pd_\mu F^{\mu\nu}$. The physical factor of the dispersion relation (\code{pdet}) is exactly $(\vect k^2-\omega^2)^2\,[a^2(\vect k^2-\omega^2)+1]^3$, with roots $\omega=\pm|\vect k|$ and $\omega=\pm\sqrt{\vect k^2+1/a^2}$. The example verifies that the profile $A_0=q(1-\ee^{-r/a})/(4\pi r)$ solves the field equations for $r>0$ and evaluates the limit $r\to0$.

\paragraph{Extension.} Add a mass term and examine how the spectrum is modified; or reduce to a first-order form with auxiliary fields.

\exheading{ex:cs}{Topologically massive electrodynamics in $2{+}1$ dimensions}{15\_chern\_simons\_2plus1.py}{research-flavoured}

\paragraph{Background.} In three spacetime dimensions a gauge-invariant mass for the photon is possible without a Higgs field. The Chern--Simons term $\tfrac\mu4\epsilon^{\mu\nu\rho}A_\mu F_{\nu\rho}$ added to Maxwell theory \cite{DeserJackiwTempleton1982a,DeserJackiwTempleton1982b} gives
\begin{equation}
\mathcal L=-\tfrac14F_{\mu\nu}F^{\mu\nu}+\tfrac\mu4\epsilon^{\mu\nu\rho}A_\mu F_{\nu\rho},\qquad
\pd_\mu F^{\mu\nu}+\tfrac\mu2\epsilon^{\nu\rho\sigma}F_{\rho\sigma}=0.
\end{equation}
The Chern--Simons term is \emph{not} gauge invariant but its variation is a total derivative: $\delta\mathcal L_{\mathrm{CS}}=\pd_\mu K^\mu$ with $K^\mu=\tfrac\mu4\epsilon^{\mu\nu\rho}\alpha F_{\nu\rho}$, so the action is invariant for gauge parameters that vanish at the boundary. (For non-abelian generalisations, ``large'' gauge transformations force the coefficient to be quantised \cite{DeserJackiwTempleton1982a}.) Defining $\dual F^\nu=\tfrac12\epsilon^{\nu\rho\sigma}F_{\rho\sigma}$, the equation is $\pd_\mu F^{\mu\nu}+\mu\dual F^\nu=0$, from which $(\Box+\mu^2)\dual F^\nu=0$: a single massive excitation of mass $|\mu|$ and spin $\sgn(\mu)$, with one degree of freedom (compared with one for the massless photon in $2{+}1$ dimensions, which is a scalar in disguise). Canonically, $N=3$ pairs, two first-class constraints ($\pi^0\weq0$ and a modified Gauss law $\pd_i\pi^i+\tfrac\mu2\epsilon^{ij}\pd_iA_j\weq0$ in which the magnetic flux acts as a source of electric charge), so $n_{\mathrm{dof}}=(6-4)/2=1$.

In the library:
\begin{lstlisting}
from fieldsmith import Model, Spacetime
from fieldsmith.hamiltonian import Hamiltonian
M = Model(Spacetime.minkowski(3))
M.parameters("mu", positive=True)
M.vector("A")
M.define(r"F_{\mu\nu}", r"\partial_\mu A_\nu - \partial_\nu A_\mu")
L = M.lagrangian(r"-\frac14 F_{\mu\nu}F^{\mu\nu} + \frac{\mu}{4}\epsilon^{\mu\nu\rho}A_\mu F_{\nu\rho}")
print(Hamiltonian(M, L).dirac_analysis().dof)             # 1
\end{lstlisting}

\paragraph{Results.} The example verifies that $\delta\mathcal L_{\mathrm{CS}}\neq0$ but $\delta\mathcal L_{\mathrm{CS}}-\pd_\mu K^\mu=0$; \code{fit} gives $\pd_\mu F^{\mu\nu}+\tfrac\mu2\epsilon^{\nu\rho\sigma}F_{\rho\sigma}$ with coefficients $\{1,\tfrac\mu2\}$; the physical dispersion factor is $\vect k^2+\mu^2-\omega^2$; and the Dirac analysis returns $3$ pairs, $2$ first-class constraints and one degree of freedom, with the secondary constraint $-\tfrac\mu2\pd_yA_1+\tfrac\mu2\pd_xA_2+\pd_x\pi^1+\pd_y\pi^2\weq0$.

\paragraph{Extension.} Couple a charged scalar and show that the Gauss-law constraint attaches magnetic flux to charge, the origin of anyonic statistics.

\exheading{ex:bpst}{The BPST instanton}{16\_bpst\_instanton.py}{research-flavoured}

\paragraph{Background.} Instantons are finite-action solutions of the Euclidean field equations. For $\mathrm{SU}(2)$ Yang--Mills theory in four Euclidean dimensions with action $S=\tfrac14\int\dd^4x\,F^a_{\mu\nu}F^a_{\mu\nu}$, the topological charge
\begin{equation}
Q=\frac{g^2}{32\pi^2}\int\dd^4x\,F^a_{\mu\nu}\dual F^a_{\mu\nu}
\end{equation}
is an integer (the winding number of the gauge transformation at infinity), and the action obeys the Bogomolny bound \cite{BPST1975}
\begin{equation}
S=\tfrac18\int\big(F^a_{\mu\nu}\mp\dual F^a_{\mu\nu}\big)^2\pm\tfrac14\int F^a_{\mu\nu}\dual F^a_{\mu\nu}\ \ge\ \frac{8\pi^2}{g^2}|Q|,
\end{equation}
saturated by (anti-)self-dual fields $F=\pm\dual F$, which automatically satisfy $D_\mu F_{\mu\nu}=0$ by the Bianchi identity. The BPST solution in regular gauge, with size $\rho$ and centred at the origin, is
\begin{equation}
A^a_\mu=\frac{2\,\eta^a_{\mu\nu}x_\nu}{g(x^2+\rho^2)},\qquad
F^a_{\mu\nu}=-\frac{4\,\eta^a_{\mu\nu}\rho^2}{g(x^2+\rho^2)^2},
\end{equation}
with the 't~Hooft symbols $\eta^a_{ij}=\epsilon_{aij}$, $\eta^a_{i4}=\delta_{ai}=-\eta^a_{4i}$. The action density is $\tfrac14F^2=48\rho^4/[g^2(x^2+\rho^2)^4]$ and $S=8\pi^2/g^2$, $Q=1$. Instantons describe tunnelling between topologically distinct vacua and underlie the $\theta$-vacuum and the $U(1)_A$ problem.

\paragraph{In the library.} The ansatz is a dictionary from the gauge-field component functions to explicit expressions; \code{M.define} builds the dual field strength.
\begin{lstlisting}
import sympy as sp
from fieldsmith import Model, Spacetime
from fieldsmith.gauge import GaugeTheory, LieAlgebra
M = Model(Spacetime.euclidean(4))
G = GaugeTheory(M)
G.add_group("A", LieAlgebra.su(2), "g")
g = M.symbols["g"]
rho = M.parameters("rho", positive=True)
X = M.coords
r2 = sum(c**2 for c in X)
def eta(a, mu, nu):                                  # 't Hooft symbol
    if mu < 3 and nu < 3: return sp.LeviCivita(a, mu, nu)
    if mu < 3 and nu == 3: return 1 if a == mu else 0
    if mu == 3 and nu < 3: return -1 if a == nu else 0
    return 0
Af = M.tensors["A"].funcs
ansatz = {Af[(a, mu)]: 2*sum(eta(a, mu, nu)*X[nu] for nu in range(4))/(g*(r2 + rho**2))
          for a in range(3) for mu in range(4)}
M.define(r"\tilde{F}^a_{\mu\nu}", r"\frac12\epsilon_{\mu\nu\rho\sigma}F^a_{\rho\sigma}")
Fa, Fd = M.tensors["F"].data, M.tensors["Ftilde"].data
print(all(sp.simplify((Fa[k] - Fd[k]).subs(ansatz).doit()) == 0 for k in Fa))   # True: self-dual
\end{lstlisting}

\paragraph{Results.} The field strength computed by the library from the ansatz is self-dual (asserted), with $F^3_{12}=-4\rho^2/[g(x^2+\rho^2)^2]$ as expected; the Euclidean field equations are satisfied exactly and symbolically (asserted); the action density equals $48\rho^4/[g^2(x^2+\rho^2)^4]$; with $\int\dd^4x=2\pi^2\int s^3\dd s$ the action is $8\pi^2/g^2$, saturating the bound, and the topological charge evaluates to $Q=1$.

\paragraph{Extension.} Check the anti-instanton, a two-instanton ansatz of 't~Hooft form, or the instanton in singular gauge.

\exheading{ex:monopole}{The 't~Hooft--Polyakov magnetic monopole}{17\_tHooft\_Polyakov\_monopole.py}{research-flavoured}

\paragraph{Background.} Consider $\mathrm{SU}(2)$ gauge theory with a Higgs field $\phi^a$ in the adjoint representation (the Georgi--Glashow model),
\begin{equation}
\mathcal L=-\tfrac14F^a_{\mu\nu}F^{a\mu\nu}+\tfrac12D_\mu\phi^aD^\mu\phi^a-V(\phi),
\end{equation}
where $V\to0$ in the BPS limit $\lambda\to0$ with $|\phi|\to v$ at infinity. The vacuum breaks $\mathrm{SU}(2)\to\mathrm U(1)$, leaving a massless photon and two $W$ bosons of mass $gv$. Because $\pi_2(\mathrm{SU}(2)/\mathrm U(1))=\pi_2(S^2)=\mathbb Z$, the theory has topologically stable, finite-energy solutions carrying magnetic charge \cite{tHooft1974,Polyakov1974}. In the hedgehog ansatz
\begin{equation}
\phi^a=\frac{x^a}{g\,r^2}H(gvr),\qquad A^a_i=\epsilon_{aij}\frac{x_j}{g\,r^2}\,[1-K(gvr)],\qquad A^a_0=0,
\end{equation}
the field equations reduce to ordinary differential equations. In the BPS limit Prasad and Sommerfield \cite{PrasadSommerfield1975} found the exact solution $K(u)=u/\sinh u$, $H(u)=u\coth u-1$. The energy $E=\int\dd^3x\,\big[\tfrac12(B^a_i)^2+\tfrac12(D_i\phi^a)^2\big]$ (with $B^a_i=-\tfrac12\epsilon_{ijk}F^a_{jk}$) obeys the Bogomolny completion
\begin{equation}
E=\tfrac12\int(B^a_i\mp D_i\phi^a)^2\pm\int B^a_iD_i\phi^a\ \ge\ v\,\Big|\!\oint\!B^a_i\hat\phi^a\,\dd S_i\Big|=\frac{4\pi v}{g},
\end{equation}
because the magnetic flux through the sphere at infinity is $4\pi/g$ (consistent with Dirac's quantisation condition for the charge $g/2$ of an $\mathrm{SU}(2)$ doublet). The solution saturates it with the first-order Bogomolny equations $B^a_i=\pm D_i\phi^a$, so the monopole mass is $M=4\pi v/g=(4\pi/g^2)M_W$. A numerical check: the library can find which sign of $A$ is a solution.

\paragraph{In the library.} The profile functions are entered explicitly and the residual of the full second-order equations is evaluated at random points to $30$ significant digits.
\begin{lstlisting}
from fieldsmith import Model
from fieldsmith.gauge import GaugeTheory, LieAlgebra
M = Model()
G = GaugeTheory(M)
G.add_group("A", LieAlgebra.su(2), "g")
M.parameters("v", positive=True)
M.field("phi", ("ad_A",), "-")                       # adjoint Higgs triplet phi^a
L = M.lagrangian(r"-\frac14 F^a_{\mu\nu}F^{a\mu\nu} + \frac12 D_\mu\phi^a D^\mu\phi^a")
E = M.euler_lagrange(L)
M.define(r"B^a_i", r"-\frac12\epsilon_{ijk}F^a_{jk}")
import sympy as sp
from fieldsmith.analysis import numeric_residual
g = M.symbols["g"]; v = M.symbols["v"]
t, x, y, z = M.coords; X = [x, y, z]
r = sp.sqrt(x**2 + y**2 + z**2); u = g*v*r
K = u/sp.sinh(u); H = u/sp.tanh(u) - 1                       # Prasad-Sommerfield profiles
Af, pf = M.tensors["A"].funcs, M.tensors["phi"].funcs
sol = {}
for a in range(3):
    sol[pf[(a,)]] = X[a]*H/(g*r**2)
    sol[Af[(a, 0)]] = 0
    for i in range(3):
        sol[Af[(a, i+1)]] = sum(sp.LeviCivita(a, i, j)*X[j] for j in range(3))*(1 - K)/(g*r**2)
print(numeric_residual(M, E.exprs, sol, n_points=2, param_values={g: sp.Rational(7, 10), v: sp.Rational(13, 10)}))
\end{lstlisting}

\paragraph{Results.} For the hedgehog with sign $s=+1$ the maximum residual is $\sim10^{-164}$ (zero to working precision), while for $s=-1$ it is $0.45$, so the correct orientation is identified. The Bogomolny equation $B^a_i=\pm D_i\phi^a$ is satisfied to the same precision for one sign and not for the other (residual $0.74$). The mass integral $\int\dd^3x\,\mathcal E$, evaluated by numerical quadrature of the radial energy density to $\infty$ with $g=v=1$, is $12.56637061=4\pi$.

\paragraph{Extension.} Switch on $\lambda>0$ and solve the radial ODEs numerically with the library's equations as the source; compare the mass to $4\pi v/g\times f(\lambda/g^2)$ with $f(0)=1$.

\exheading{ex:symmetries}{A tour of spacetime symmetries and their Noether currents}{18\_symmetries\_tour.py}{intermediate--advanced}

\paragraph{Background.} Noether's theorem \cite{Noether1918} associates a conserved quantity with every continuous symmetry. The Poincar\'e group of Minkowski space gives the full set of ``mechanical'' conservation laws: translations ($T^{\mu\nu}$ and the four-momentum), rotations (angular momentum $J^{ij}=\int(x^iT^{0j}-x^jT^{0i})$) and boosts ($K^i=\int(x^iT^{00}-tT^{0i})$, whose conservation says that the centre of energy moves uniformly). Scale transformations $x\to\lambda x$, $\phi\to\lambda^{-1}\phi$ (in four dimensions) are a symmetry of massless $\phi^4$ theory; the dilatation current is $J^\mu=-(x^\nu T^\mu{}_\nu+\phi\pd^\mu\phi)$, and on shell
\begin{equation}
\pd_\mu J^\mu=\phi\,V'(\phi)-4V(\phi),
\end{equation}
which vanishes if and only if $V\propto\phi^4$. With a mass term, the divergence is $-m^2\phi^2$: scale invariance is broken by exactly the mass term. A pure shift $\phi\to\phi+c$ is a symmetry of the massless free field, with current $J^\mu=\pd^\mu\phi$; any potential breaks it and the defect is $-V'(\phi)$.

\paragraph{In the library.} A symmetry is a dictionary of variations (for spacetime symmetries, the transport term $-\xi^\mu\pd_\mu\phi$) together with the boundary term $K^\mu$ that makes $\delta\mathcal L=\pd_\mu K^\mu$.
\begin{lstlisting}
import sympy as sp
from fieldsmith import Model
from fieldsmith.noether import noether_current, symmetry_defect
M = Model()
M.parameters("m lambda", positive=True)
phi = M.scalar("phi")
t, x, y, z = X = M.coords
L = M.lagrangian(r"\frac12\partial_\mu\phi\partial^\mu\phi - \frac12 m^2\phi^2 - \frac{\lambda}{4!}\phi^4")
dx = [x, t, 0, 0]                                           # boost generator  delta x^mu
var = {phi: -sum(dx[m]*sp.diff(phi, X[m]) for m in range(4))}
K = [-dx[m]*L for m in range(4)]                            # delta L = d_mu K^mu
J = noether_current(M, L, var, K)                           # checks the symmetry first
print(J.identity_holds())                                   # True
\end{lstlisting}

\paragraph{Results.} For the massive $\phi^4$ Lagrangian, all four translation currents (the rows of the canonical stress tensor), the boost along $x$, and the rotation about $z$ pass the symmetry check and satisfy the Noether identity; the boost charge density is $J^0=-(xT^{00}-tT^{01})$ (printed in components; the overall sign follows from the sign convention of the transport term) and the rotation charge is $J^0=(y\pd_x\phi-x\pd_y\phi)\dot\phi=xT^{02}-yT^{01}$. The shift current of the free massless field is $J^\mu=\pd^\mu\phi$; for the interacting theory \code{noether\_current} raises ``not a symmetry'' and reports the defect $-V'(\phi)$. The dilatation defect is $0$ for the massless quartic theory and $-m^2\phi^2$ with a mass term, as predicted.

\paragraph{Extension.} Find the special conformal currents of the massless field (with the improved tensor of \cref{ex:stress}); verify that without improvement the symmetry check fails.

\section{Discussion}
\label{sec:discussion}

\subsection{Limitations}
\label{sec:limits}

We list the limitations of the current version as precisely as we can; each is documented in the repository and, where possible, enforced by an explicit error.

\begin{description}[leftmargin=1.2em]
\item[Component-based cost.] Because tensors are stored by components and SymPy simplifies the resulting expressions, the cost grows with the spacetime dimension $d$ and the dimension of the gauge group. $\mathrm{SU}(2)$ and electroweak calculations take seconds to tens of seconds; the $\mathrm{SU}(3)$ Lagrangian has $1650$ terms--construction and the Bianchi identity are feasible (the whole Yang--Mills example runs in under a minute) but heavier analyses will be slow; $\mathrm{SU}(N)$ with $N\ge4$ is impractical.
\item[Classical, commuting fields.] The Dirac field is a classical commuting spinor. No Grassmann variables, hence no classical analysis of fermionic symmetries such as supersymmetry, and no ghosts.
\item[Canonical module.] The Hamiltonian module requires at most first derivatives in time and Lagrangians that are (at most) quadratic in velocities with an algebraically invertible Hessian. Higher-derivative theories (Podolsky) can be studied at the level of field equations, Noether currents and dispersion, but not canonically; the Ostrogradsky construction is not implemented. Born--Infeld-type Lagrangians are refused. The Dirac--Bergmann implementation is heuristic in two respects: the weak equality is imposed by pivoting on a momentum derivative, and the classification into first and second class uses non-vanishing brackets modulo total derivatives; it is validated on the textbook cases in this paper but is not a general-purpose constraint solver.
\item[Hilbert tensor.] The metric variation is first-order about flat space, takes the Lagrangian as a string and supports bosonic fields only (no tetrad).
\item[Gravity.] Curvature tensors, covariant derivatives and the Einstein tensor are provided for a given metric, but the metric is not dynamical: there is no Einstein--Hilbert variation and no linearised-gravity module.
\item[Noether currents on curved backgrounds.] The Noether module is validated in flat spacetime. On curved backgrounds the identity uses the density $\sqrt{|g|}$ but symmetries must be supplied as explicit variations (Killing vectors are not computed).
\item[Gauge fixing and BRST.] There is no gauge-fixing, ghost, or BRST machinery; gauge fixing terms may be added by hand to the Lagrangian.
\item[Parser.] No \verb|\sum| or \verb|\int|; derivative operators act on the next postfix term; numeric adjoint indices are $1$-based whereas spacetime numerics are $0$-based; gamma matrices exist for the $4$D Dirac representation only. Some \LaTeX{} constructs (e.g.\ \verb|\left|, \verb|\right|, \verb|\text|) are silently dropped.
\item[Dispersion analysis.] The gcd-of-minors method yields the divisorial part of the rank-drop locus and depends on coefficients being constant at the wave-vector of interest; non-constant backgrounds require an explicit local treatment.
\end{description}

\subsection{Related software}
\label{sec:related}

Computer algebra for field theory is a mature area; we describe where \fs{} sits.
\begin{itemize}
\item \textbf{General symbolic tensor algebra.} \textsc{xAct} (\textsc{Mathematica}) and \textsc{Cadabra} \cite{Peeters2007} provide abstract-index tensor canonicalisation and manipulation, and are the tools of choice for large tensor calculations such as higher-order perturbation theory in general relativity. SymPy itself offers \code{sympy.tensor} and \code{sympy.diffgeom}. \fs{} works with explicit components instead of abstract indices; this makes it less scalable but removes the need for a canonicaliser and gives a direct handle on every number.
\item \textbf{Amplitude and loop calculations.} \textsc{FORM} \cite{Vermaseren2000} and \textsc{FeynCalc} \cite{Mertig1991,Shtabovenko2016} excel at Dirac and colour algebra for Feynman diagrams. \fs{} does not compute amplitudes.
\item \textbf{Variational calculus.} SymPy provides \code{euler\_equations}, which \fs{} uses; the novelty here is the surrounding structure (index notation, Noether, constraints, gauge theory) and the \LaTeX{} front end.
\item \textbf{Teaching tools.} Computer-algebra worksheets for classical field theory exist in individual courses; to our knowledge, there is no open-source package that combines a \LaTeX{}-index interface, a verified Noether/Dirac pipeline and an auto-grader.
\end{itemize}
We emphasise that these tools are complementary: for example, a result derived in \fs{} for a small gauge group can be cross-checked in \textsc{xAct} at the level of abstract expressions.

\subsection{Future directions}

The following extensions seem natural and are of increasing difficulty:
(i) a more complete \LaTeX{} grammar (explicit sums and integrals, bracketed indices, \verb|\mathcal|, \verb|\bm|); (ii) a unified treatment of non-abelian vector theories with Chern--Simons terms and Wilson lines; (iii) an Ostrogradsky module with automatic recognition of ghost modes for higher-derivative actions; (iv) dynamical gravity: linearised Einstein--Hilbert, the Fierz--Pauli theory and its degree-of-freedom analysis, and eventually the ADM decomposition; (v) a Grassmann extension for fermionic symmetry analysis; (vi) a BRST module; (vii) differential forms in place of components, which would make Chern--Simons and Wess--Zumino terms natural; (viii) performance work, including component sparsity, caching of expansions and optional use of a faster back-end. Contributions are welcome.

\subsection{Conclusion}

We have presented \fs{}, a compact library that turns a \LaTeX{} Lagrangian into verified classical field theory: equations of motion, symmetries, stress tensors, constraint analysis, spectra and gauge-theoretic data. We described its theoretical underpinnings, its implementation and the checks that guard it, and we illustrated its use on eighteen examples that run from the Klein--Gordon field to the 't~Hooft--Polyakov monopole. The documentation, the examples and this paper together aim at a single purpose: to make it easy to derive existing theory and to explore new theory with confidence that the algebra is right. The physics in the examples is standard; the contribution is a tool that makes it reproducible, inspectable and quick to modify.

\section*{Acknowledgements}
The author thanks Konstantin V. Stepanyants (Moscow State University) whose extensive academic lectures and educational video series inspired the creation of the library. The library builds on \textsc{SymPy} and its community.

\section*{Code and data availability}
The source code, examples, tests and documentation are available at \url{https://github.com/nikola-m/fieldsmith} under the MIT licence. All numbers in this paper can be regenerated with the scripts described in \cref{app:repro}.

\appendix
\section{Conventions}
\label{app:conventions}

\begin{table}[H]
\centering\small
\begin{tabular}{@{}p{4.2cm}p{10.3cm}@{}}
\toprule
Item & Convention \\
\midrule
Units & $\hbar=c=1$ \\
Signature & $(+,-,-,-)$ by default; \code{Spacetime.minkowski(d, "mostly\_plus")} for $(-,+,\dots,+)$; \code{Spacetime.euclidean(d)} for $\delta_{\mu\nu}$ \\
Coordinates & $x^0=t$; spacetime component indices are $0$-based \\
Levi-Civita & $\epsilon^{0123}=+1/\sqrt{|g|}$ (so $\epsilon_{0123}=-1$ in Minkowski space); \code{Model(eps0123=-1)} flips the sign; spatial $\epsilon_{ijk}$ with $\epsilon_{123}=1$ \\
Index kinds & Greek: spacetime. Latin $i,j,k,l,m,n$: spatial ($1,\dots,d-1$). Latin $a$--$f$: adjoint of the first gauge factor (and further families for further factors). User-defined: any letters registered with \code{add\_index\_space} \\
Adjoint numerals & $1$-based (\verb|A^1_\mu|, \dots, \verb|A^3_\mu|) \\
Covariant derivative & $D_\mu=\pd_\mu-\ii gA^a_\mu T^a$ on a multiplet; $D_\mu\chi^a=\pd_\mu\chi^a+gf^{abc}A^b_\mu\chi^c$ on an adjoint \\
Field strength & $F^a_{\mu\nu}=\pd_\mu A^a_\nu-\pd_\nu A^a_\mu+gf^{abc}A^b_\mu A^c_\nu$ \\
Gauge transformation & $\delta\phi=\ii\alpha^aT^a\phi$, $\delta A^a_\mu=g^{-1}\pd_\mu\alpha^a+f^{abc}A^b_\mu\alpha^c$ \\
Generators & $\mathrm{SU}(N)$: generalised Gell-Mann matrices, $\tr(T^aT^b)=\tfrac12\delta^{ab}$ \\
Electric/magnetic fields & $E_i=F_{i0}$, $B_i=\tfrac12\epsilon_{ijk}F_{jk}$ (overall sign opposite to the textbook vector fields; irrelevant for quadratic forms) \\
Riemann tensor & MTW \cite{MTW}: $R^\rho{}_{\sigma\mu\nu}=\pd_\mu\Gamma^\rho_{\nu\sigma}-\pd_\nu\Gamma^\rho_{\mu\sigma}+\Gamma^\rho_{\mu\lambda}\Gamma^\lambda_{\nu\sigma}-\Gamma^\rho_{\nu\lambda}\Gamma^\lambda_{\mu\sigma}$ \\
Stress tensor & Hilbert: $T_{\mu\nu}=(2/\sqrt{|g|})\,\delta(\sqrt{|g|}\mathcal L)/\delta g^{\mu\nu}$ for $(+,-,-,-)$ \\
Complex fields & $\phi$ and $\phi^*$ independent; $\delta S/\delta\phi^*$ is the equation of motion of $\phi$ \\
Dirac matrices & Dirac representation; $\bar\psi=\psi^\dagger\gamma^0$ \\
EL sign & the EL expression is $\pd\mathcal L/\pd\varphi-\pd_\mu\pd\mathcal L/\pd(\pd_\mu\varphi)+\dots$ \\
\bottomrule
\end{tabular}
\caption{Conventions used by the library and in this paper.}
\end{table}

\section{The \LaTeX{} subset}
\label{app:grammar}

The grammar accepted by the parser is, informally,
\begingroup\small\begin{align*}
\mathit{expr}&::=\mathit{term}\ (\,(\texttt{+}\,|\,\texttt{-})\ \mathit{term}\,)^*\\
\mathit{term}&::=\mathit{factor}\ (\,(\texttt{*}\,|\,\texttt{\textbackslash cdot}\,|\,\texttt{\textbackslash times}\,|\,\texttt{/}\,|\,\text{implicit})\ \mathit{factor}\,)^*\\
\mathit{factor}&::=(\texttt{-}\,|\,\texttt{+})\ \mathit{factor}\ |\ \mathit{postfix}\\
\mathit{postfix}&::=\mathit{operator}^*\ \mathit{atom}\ (\,\texttt{\^{}}\mathit{exponent}\,|\,\texttt{\^{}\textbackslash dagger}\,|\,\texttt{\^{}*}\,|\,\texttt{!}\,)^*\\
\mathit{operator}&::=\texttt{\textbackslash partial\_}\mu\ |\ \texttt{\textbackslash nabla\_}\mu\ |\ \texttt{\textbackslash Box}\ |\ \texttt{D\_}\mu\ |\ \texttt{\textbackslash bar}\ |\ \text{registered operator}\\
\mathit{atom}&::=\text{number}\ |\ \texttt{\textbackslash frac\{..\}\{..\}}\ |\ \texttt{\textbackslash sqrt\{..\}}\ |\ \text{function}\texttt{(..)}\ |\ \texttt{(}\mathit{expr}\texttt{)}\ |\ \texttt{\{}\mathit{expr}\texttt{\}}\ |\ \mathit{symbol}\,[\text{indices}]
\end{align*}\endgroup
Supported functions: \texttt{\textbackslash sin, \textbackslash cos, \textbackslash tan, \textbackslash sinh, \textbackslash cosh, \textbackslash tanh, \textbackslash exp, \textbackslash ln, \textbackslash log, \textbackslash arcsin, \textbackslash arccos, \textbackslash arctan, \textbackslash sech, \textbackslash cot}. Spacing commands (\texttt{\textbackslash ,}, \texttt{\textbackslash ;}, \texttt{\textbackslash quad}, \dots) and \texttt{\textbackslash left}, \texttt{\textbackslash right}, \texttt{\textbackslash mathrm}, \texttt{\textbackslash mathbf}, \texttt{\textbackslash text}, \texttt{\textbackslash displaystyle} are ignored. Greek letters, accents (\texttt{\textbackslash tilde}, \texttt{\textbackslash hat}, \texttt{\textbackslash dot}) and primes are part of symbol names.

\paragraph{Semantics.} (1) A repeated index label within a product is summed over its range. (2) An operator applies to the next \emph{postfix} item; use parentheses to widen the scope. (3) Superscripts on a symbol with a registered index structure are indices; otherwise they are powers. For example, in \verb|F^{\mu\nu}| the braces hold indices, in \verb|\phi^2| the 2 is a power; \verb|(F_{\mu\nu})^2| is the self-contraction $F_{\mu\nu}F^{\mu\nu}$. (4) Subscripts and superscripts that are single Greek letters or label letters are indices when the head symbol is a field or tensor; for parameters they are part of the name (e.g.\ \verb|m_a|, \verb|c_1|). (5) \verb|g| denotes the metric when followed by indices and the coupling parameter otherwise.

\section{API summary}
\label{app:api}

\begin{table}[H]
\centering\small
\begin{tabular}{@{}p{5.6cm}p{9.0cm}@{}}
\toprule
Call & Result \\
\midrule
\code{Model(spacetime, strict, eps0123)} & model on a given spacetime (default $3{+}1$ Minkowski) \\
\code{M.parameters("m g", **assume)} & symbols \\
\code{M.scalar / vector / tensor / multiplet / field} & declare fields \\
\code{M.add\_index\_space(kind, letters, dim)} & custom index family \\
\code{M.define(lhs, rhs)} & derived tensor, usable in later strings \\
\code{M.lagrangian(s)} & parse and expand a scalar Lagrangian \\
\code{M.parse(s, as\_tensor)} & parse arbitrary \LaTeX{} expression \\
\code{M.check(s)} & verify an identity \texttt{lhs = rhs} \\
\code{M.euler\_lagrange(L)} & \code{EquationSet}: EL equations (any order) \\
\code{E.tensor(name)} / \code{E[name]} & field equations as tensor / expression \\
\code{E.check\_solution(map, numeric)} & test a candidate solution \\
\code{fit(M, tensor, basis)} & coefficients of covariant decomposition, or \code{None} \\
\code{noether\_current(M, L, var, K)} & \code{NoetherCurrent}: $J^\mu$, \code{identity\_holds()} \\
\code{symmetry\_defect(M, L, var, K)} & $\delta\mathcal L-\pd_\mu K^\mu$ \\
\code{complex\_variation(M, name, expr)} & variation of a complex field and its conjugate \\
\code{canonical\_stress\_tensor(M, L)} & $T^\mu{}_\nu$ \\
\code{hilbert\_stress\_tensor(M, L\_str)} & $T_{\mu\nu}$ (flat, string Lagrangian) \\
\code{Hamiltonian(M, L)} & $\pi$, $H$, \code{bracket}, \code{hamilton\_equations}, \code{dirac\_analysis} \\
\code{linearize(M, E, background)} & linear equations and perturbation fields \\
\code{dispersion(M, exprs, funcs)} & $K$, $\det K$, $\omega$-roots, rank, \code{pdet} \\
\code{LieAlgebra.su(N) / u1()} & generators, $f^{abc}$, Casimir, Jacobi test \\
\code{GaugeTheory(M)} & \code{add\_group}, \code{add\_matter}, \code{gauge\_parameters}, \code{gauge\_variation}, \code{gauge\_mass\_matrix} \\
\code{scalar\_mass\_matrix(M, V, fields, vac)} & Hessian of a potential \\
\code{grade\_field\_equation / grade\_identity / grade\_conserved\_current} & truthy \code{Grade} with a message \\
\bottomrule
\end{tabular}
\caption{Principal entry points of \fs{}.}
\end{table}

\section{Reproducing the results}
\label{app:repro}

Every listing in \cref{part:examples} was executed against the library, and the output quoted is the printed output (the files \texttt{paper/snippets/s*.py} contain the listings and \texttt{paper/snippets/s*.out} the captured output). The full example scripts are in \texttt{examples/}, and \texttt{tools/build\_gallery.py} regenerates \texttt{docs/gallery.md} with the complete output of all eighteen. The tests are run by
\begin{lstlisting}[language=bash,frame=single]
pip install -e ".[test]"
python -m pytest -m "not slow"     # 53 unit/integration tests (~1 min)
python -m pytest                    # plus the 18 example scripts (~5 min)
\end{lstlisting}
\Cref{tab:timings} gives per-example wall-clock times in the environment used to prepare this paper (a single core of a cloud virtual machine, Python~3 with SymPy~1.14); absolute figures will vary with hardware and SymPy version.
\begin{table}[H]
\centering\small
\begin{tabular}{@{}rlr@{}}
\toprule
\# & Example & Time (s) \\
\midrule
01 & Klein--Gordon & 1.0 \\
02 & Kink, sine-Gordon & 12.2 \\
03 & Maxwell & 12.1 \\
04 & Proca, St\"uckelberg & 16.4 \\
05 & Abelian Higgs & 24.6 \\
06 & O($N$) Goldstone & 3.7 \\
07 & Yang--Mills SU(2), SU(3) & 44.3 \\
08 & Electroweak & 15.2 \\
09 & QED, Dirac & 8.5 \\
10 & Stress tensors & 9.5 \\
11 & Curved spacetime & 4.4 \\
12 & Axion electrodynamics & 5.4 \\
13 & Vacuum birefringence & 47.0 \\
14 & Podolsky & 62.2 \\
15 & Chern--Simons (2+1) & 2.6 \\
16 & BPST instanton & 10.3 \\
17 & 't Hooft--Polyakov & 27.9 \\
18 & Symmetries tour & 2.0 \\
\midrule
 & total & 309\\
\bottomrule
\end{tabular}
\caption{Wall-clock time for each example script.}
\label{tab:timings}
\end{table}

\bibliographystyle{unsrtnat}
\bibliography{refs}

@article{Peeters2007,
  author={Peeters, Kasper}, title={Cadabra: a field-theory motivated symbolic computer algebra system},
  journal={Comput. Phys. Commun.}, volume={176}, pages={550--558}, year={2007}}

@misc{Vermaseren2000,
  author={Vermaseren, J. A. M.}, title={New features of {FORM}}, year={2000}, eprint={math-ph/0010025}}

@article{Mertig1991,
  author={Mertig, R. and B{\"o}hm, M. and Denner, A.}, title={{FeynCalc} -- computer-algebra calculations for {Feynman} amplitudes},
  journal={Comput. Phys. Commun.}, volume={64}, pages={345--359}, year={1991}}

@article{Shtabovenko2016,
  author={Shtabovenko, V. and Mertig, R. and Orellana, F.}, title={New developments in {FeynCalc} 9.0},
  journal={Comput. Phys. Commun.}, volume={207}, pages={432--444}, year={2016}}

@article{Meurer2017,
  author={Meurer, Aaron and others}, title={{SymPy}: symbolic computing in {Python}},
  journal={PeerJ Computer Science}, volume={3}, pages={e103}, year={2017}}

@book{PeskinSchroeder,
  author={Peskin, Michael E. and Schroeder, Daniel V.}, title={An Introduction to Quantum Field Theory},
  publisher={Addison-Wesley}, year={1995}}

@book{LandauLifshitz,
  author={Landau, L. D. and Lifshitz, E. M.}, title={The Classical Theory of Fields}, edition={4th},
  publisher={Pergamon}, year={1975}}

@book{WeinbergQFT1,
  author={Weinberg, Steven}, title={The Quantum Theory of Fields, Volume I: Foundations},
  publisher={Cambridge University Press}, year={1995}}

@book{WeinbergQFT2,
  author={Weinberg, Steven}, title={The Quantum Theory of Fields, Volume II: Modern Applications},
  publisher={Cambridge University Press}, year={1996}}

@book{HenneauxTeitelboim,
  author={Henneaux, Marc and Teitelboim, Claudio}, title={Quantization of Gauge Systems},
  publisher={Princeton University Press}, year={1992}}

@book{MTW,
  author={Misner, Charles W. and Thorne, Kip S. and Wheeler, John A.}, title={Gravitation},
  publisher={W. H. Freeman}, year={1973}}

@book{Rajaraman,
  author={Rajaraman, R.}, title={Solitons and Instantons}, publisher={North-Holland}, year={1982}}

@article{Dirac1950,
  author={Dirac, P. A. M.}, title={Generalized {Hamiltonian} dynamics},
  journal={Can. J. Math.}, volume={2}, pages={129--148}, year={1950}}

@article{Ostrogradsky1850,
  author={Ostrogradsky, M.}, title={M{\'e}moires sur les {\'e}quations diff{\'e}rentielles relatives au probl{\`e}me des isop{\'e}rim{\`e}tres},
  journal={Mem. Ac. St. Petersbourg}, volume={VI}, number={4}, pages={385}, year={1850}}

@incollection{Woodard2007,
  author={Woodard, Richard P.}, title={Avoiding dark energy with 1/{R} modifications of gravity},
  booktitle={The Invisible Universe: Dark Matter and Dark Energy}, series={Lect. Notes Phys.}, volume={720},
  pages={403--433}, year={2007}}

@article{Bogomolny1976,
  author={Bogomol'nyi, E. B.}, title={The stability of classical solutions},
  journal={Sov. J. Nucl. Phys.}, volume={24}, pages={449--454}, year={1976}}

@article{Proca1936,
  author={Proca, A.}, title={Sur la th{\'e}orie ondulatoire des {\'e}lectrons positifs et n{\'e}gatifs},
  journal={J. Phys. Radium}, volume={7}, pages={347--353}, year={1936}}

@article{Stueckelberg1938,
  author={St{\"u}ckelberg, E. C. G.}, title={Die Wechselwirkungskr{\"a}fte in der Elektrodynamik und in der Feldtheorie der Kernkr{\"a}fte},
  journal={Helv. Phys. Acta}, volume={11}, pages={225--244}, year={1938}}

@article{Higgs1964,
  author={Higgs, Peter W.}, title={Broken symmetries and the masses of gauge bosons},
  journal={Phys. Rev. Lett.}, volume={13}, pages={508--509}, year={1964}}

@article{EnglertBrout1964,
  author={Englert, F. and Brout, R.}, title={Broken symmetry and the mass of gauge vector mesons},
  journal={Phys. Rev. Lett.}, volume={13}, pages={321--323}, year={1964}}

@article{GHK1964,
  author={Guralnik, G. S. and Hagen, C. R. and Kibble, T. W. B.}, title={Global conservation laws and massless particles},
  journal={Phys. Rev. Lett.}, volume={13}, pages={585--587}, year={1964}}

@article{Goldstone1961,
  author={Goldstone, J.}, title={Field theories with ``superconductor'' solutions},
  journal={Nuovo Cimento}, volume={19}, pages={154--164}, year={1961}}

@article{YangMills1954,
  author={Yang, C. N. and Mills, R. L.}, title={Conservation of isotopic spin and isotopic gauge invariance},
  journal={Phys. Rev.}, volume={96}, pages={191--195}, year={1954}}

@article{Glashow1961,
  author={Glashow, Sheldon L.}, title={Partial-symmetries of weak interactions},
  journal={Nucl. Phys.}, volume={22}, pages={579--588}, year={1961}}

@article{Weinberg1967,
  author={Weinberg, Steven}, title={A model of leptons},
  journal={Phys. Rev. Lett.}, volume={19}, pages={1264--1266}, year={1967}}

@incollection{Salam1968,
  author={Salam, Abdus}, title={Weak and electromagnetic interactions},
  booktitle={Elementary Particle Theory (Nobel Symposium No.~8)}, editor={Svartholm, N.},
  publisher={Almqvist \& Wiksell}, address={Stockholm}, pages={367--377}, year={1968}}

@article{Belinfante1940,
  author={Belinfante, F. J.}, title={On the current and the density of the electric charge, the energy, the linear momentum and the angular momentum of arbitrary fields},
  journal={Physica}, volume={7}, pages={449--474}, year={1940}}

@article{Rosenfeld1940,
  author={Rosenfeld, L.}, title={Sur le tenseur d'impulsion-{\'e}nergie},
  journal={M{\'e}m. Acad. Roy. Belg. Sci.}, volume={18}, number={6}, year={1940}}

@article{CallanColemanJackiw1970,
  author={Callan, Curtis G. and Coleman, Sidney and Jackiw, Roman}, title={A new improved energy-momentum tensor},
  journal={Ann. Phys.}, volume={59}, pages={42--73}, year={1970}}

@article{Wilczek1987,
  author={Wilczek, Frank}, title={Two applications of axion electrodynamics},
  journal={Phys. Rev. Lett.}, volume={58}, pages={1799--1802}, year={1987}}

@article{CarrollFieldJackiw1990,
  author={Carroll, Sean M. and Field, George B. and Jackiw, Roman}, title={Limits on a {Lorentz}- and parity-violating modification of electrodynamics},
  journal={Phys. Rev. D}, volume={41}, pages={1231--1240}, year={1990}}

@article{EulerHeisenberg1936,
  author={Euler, H. and Heisenberg, W.}, title={Folgerungen aus der {Diracschen} Theorie des {Positrons}},
  journal={Z. Phys.}, volume={98}, pages={714--732}, year={1936}}

@article{BialynickaBirula1970,
  author={Bialynicka-Birula, Z. and Bialynicki-Birula, I.}, title={Nonlinear effects in quantum electrodynamics. {Photon} propagation and photon splitting in an external field},
  journal={Phys. Rev. D}, volume={2}, pages={2341--2345}, year={1970}}

@article{Adler1971,
  author={Adler, Stephen L.}, title={Photon splitting and photon dispersion in a strong magnetic field},
  journal={Ann. Phys.}, volume={67}, pages={599--647}, year={1971}}

@article{Podolsky1942,
  author={Podolsky, Boris}, title={A generalized electrodynamics. {Part I} -- non-quantum},
  journal={Phys. Rev.}, volume={62}, pages={68--71}, year={1942}}

@article{DeserJackiwTempleton1982a,
  author={Deser, S. and Jackiw, R. and Templeton, S.}, title={Three-dimensional massive gauge theories},
  journal={Phys. Rev. Lett.}, volume={48}, pages={975--978}, year={1982}}

@article{DeserJackiwTempleton1982b,
  author={Deser, S. and Jackiw, R. and Templeton, S.}, title={Topologically massive gauge theories},
  journal={Ann. Phys.}, volume={140}, pages={372--411}, year={1982}}

@article{BPST1975,
  author={Belavin, A. A. and Polyakov, A. M. and Schwartz, A. S. and Tyupkin, Yu. S.}, title={Pseudoparticle solutions of the {Yang--Mills} equations},
  journal={Phys. Lett. B}, volume={59}, pages={85--87}, year={1975}}

@article{tHooft1974,
  author={'t Hooft, G.}, title={Magnetic monopoles in unified gauge theories},
  journal={Nucl. Phys. B}, volume={79}, pages={276--284}, year={1974}}

@article{Polyakov1974,
  author={Polyakov, A. M.}, title={Particle spectrum in quantum field theory},
  journal={JETP Lett.}, volume={20}, pages={194--195}, year={1974}}

@article{PrasadSommerfield1975,
  author={Prasad, M. K. and Sommerfield, Charles M.}, title={Exact classical solution for the 't {Hooft} monopole and the {Julia--Zee} dyon},
  journal={Phys. Rev. Lett.}, volume={35}, pages={760--762}, year={1975}}

@article{Noether1918,
  author={Noether, Emmy}, title={Invariante {V}ariationsprobleme},
  journal={Nachr. Ges. Wiss. G{\"o}ttingen, Math.-Phys. Kl.}, pages={235--257}, year={1918}}
\end{document}